\documentclass[twocolumn,showpacs,prb,amsmath,amssymb,floatfix,eqsecnum]{revtex4}
\usepackage{amsmath,amssymb}
\usepackage{bm}
\usepackage{epsfig}
\usepackage{psfrag}

\newcommand{\bd}{\bm}

\begin{document}

 \title{Fermi surface renormalization and confinement in two coupled metallic chains}

 \author{Sascha Ledowski and Peter Kopietz}
  
  \affiliation{Institut f\"{u}r Theoretische Physik, Universit\"{a}t
    Frankfurt,  Max-von-Laue Strasse 1, 60438 Frankfurt, Germany}

 \date{\today}
\date{August 4, 2006}

 \begin{abstract}
Using  a non-perturbative functional  
renormalization group approach involving both fermionic and bosonic fields 
we calculate the interaction-induced change of the Fermi surface
of spinless fermions moving on two chains 
connected by weak interchain hopping
$t_{\bot}$.
For a model containing interband backward scattering only we show that
the distance $ \Delta$ between the Fermi momenta
 associated with the bonding  and
the antibonding band can
be strongly reduced, corresponding to a large reduction of the
effective interchain hopping $t_{\bot}^{\ast} \propto \Delta $.
A self-consistent one-loop approximation 
neglecting marginal vertex corrections and
wave-function renormalizations predicts 
a confinement transition
for sufficiently large interchain backscattering, where
the renormalized $t_{\bot}^{\ast}$ vanishes.
However,
a more accurate calculation taking
vertex corrections and wave-function renormalizations
into account predicts only weak confinement
in the sense that  $0< | t_{\bot}^{\ast} | \ll | t_{\bot} |$.
Our method can be applied to other strong-coupling problems 
where the  dominant scattering channel is known.

\end{abstract}


 %\preprint{}

  %\draft

  \maketitle

\section{Introduction}

In strongly correlated Fermi systems
electron-electron interactions
can have  drastic effects
on the geometry and the topology of the Fermi surface.
For example, strong
forward scattering can give rise to a Pomeranchuk instability,
where  the shape of the Fermi surface spontaneously changes such that
it has a lower symmetry than
the underlying lattice \cite{Pomenanchuk58}.
Another example is the  
Lifshitz transition~\cite{Lifshitz60}, where
the topology of the
Fermi surface changes discontinuously without symmetry breaking
as a function of some
external control parameter. This   
gives rise to anomalies  in thermodynamic and kinetic 
properties of a metal.
Conditions on the range and the strength of the interaction leading to
Pomeranchuk and Lifshitz transitions have recently been derived
in Ref.~[\onlinecite{Quintanilla06}]. 

In this work we shall focus on another type of phase transition 
associated with the geometry of the Fermi surface, which we call
{\it{confinement transition}}. 
This quantum phase transition can
occur in quasi one-dimensional metals with an open Fermi surface,
consisting of two disconnected weakly curved sheets.  
Due to strong interactions, the curvature of the Fermi sheets
can be smoothed out and can eventually vanish in certain sectors.
In the extreme case, the 
renormalized Fermi surface
consists of two completely flat parallel planes.
The motion of the fermions in real space is then 
strictly one-dimensional, although
in the absence of interactions it is not.
We therefore call such a transition {\it{confinement transition}}.
In the confined state the low-energy properties of the system
resemble that of a one-dimensional Luttinger liquid.
Because the Fermi surface in the confined state
has an additional nesting symmetry, at the confinement transition  the symmetry
of the Fermi surface increases, 
in contrast to the Pomeranchuk instability, where interactions
lower the symmetry of the  Fermi surface.
An interaction-induced flattening of the Fermi surface
might  also play a role in
the Hubbard model close to half filling, where the bare Fermi surface
consists of four almost flat sectors.
Completely flat parts
of the Fermi surface can give rise to non-Fermi liquid 
behavior \cite{Luther94,Zheleznyak97,Ferraz03}.
Evidence of an interaction-induced flattening of the Fermi surface of the
Hubbard model close to half filling has been found in Ref.~[\onlinecite{Honerkamp01}].

Similar to the Pomeranchuk transition,
the  confinement transition is a strong-coupling phenomenon.
Hence, the usual weak coupling perturbative expansions  are not sufficient to study
this transition.
Due to a lack of controlled methods to deal with strongly interacting fermions
in dimensions larger than one, it is very difficult to study
the confinement transition.
To shed some light on the underlying mechanism,  we shall
in this work consider the simpler problem of just two metallic spinless chains
coupled by weak interchain hopping $t_{\bot}$.
The confined state corresponds to a vanishing
renormalized interchain hopping $ t_{\bot}^{\ast} =0$, so that
electrons cannot move from one chain to the other, in spite of the fact
that the bare interchain hopping is finite.
In a subsequent paper~\cite{Ledowski06}, we shall discuss the more difficult confinement 
problem in an infinite array of coupled chains. It turns out that the basic
mechanism
responsible for the tendency  towards confinement can
already be understood from the simpler two-chain problem. 

Because perturbation theory in the two-chain problem  
is plagued by the usual infrared divergencies
of one-dimensional Fermi systems, even in the limit of
weak interactions the Fermi surface cannot be calculated
within renormalized 
perturbation theory\cite{Neumayr03,Dusuel03}.
In dimensions $ D \geq 2$ the Fermi surface deformation has been studied
to all orders in perturbation theory in Ref.~[\onlinecite{Feldman96}]
for a general class of models.
Within the framework
of the renormalization group (RG)
the Fermi surface can be defined non-perturbatively from the
requirement that the relevant coupling constants $r_l ( {\bd{k}}_F )$ related to the
self-energy $ \Sigma ( {\bd{k}}_F ,  \omega =0)$ at the true Fermi surface 
${\bd{k}}_F$
flow into a fixed point of the RG~\cite{Kopietz01,Ledowski03}.
In Ref.~[\onlinecite{Ledowski05}] 
we have calculated  the shift of the
Fermi surface in the two-chain system
within the usual weak coupling
expansion of the  RG $\beta$-functions.
We have shown that interchain backscattering gives rise to the dominant 
logarithmic renormalization
of the distance
 $
 \Delta = k^{+} - k^{-}
 $
between the Fermi momenta $k^{+}$ and $k^{-}$ 
associated with  the bonding and the antibonding band.
Denoting by $\Delta_1$ the value of $\Delta$ within the Hartree-Fock
approximation, the self-consistency condition for the
true Fermi point distance in the spinless two-chain system
can be cast into the form
 \begin{equation}
 \Delta = \frac{\Delta_1 }{ 1 + 2 g_0^2 \ln ( \Lambda_0 / \Delta )}
 \label{eq:deltaprevious}
 \; ,
 \end{equation}
where $\Lambda_0$ is an ultraviolet cutoff and
$g_0$ is the bare value of the dimensionless coupling constant describing
interchain backscattering, which will be defined more precisely in
Sec.~\ref{subsec:relevant}.
From Eq.~(\ref{eq:deltaprevious})
we see that sufficiently large interchain backscattering strongly reduces
the value of $\Delta$. But the renormalized $\Delta$ never vanishes, so that
there is no true confinement transition. One should keep in mind, however, that
Eq.~(\ref{eq:deltaprevious}) has been derived by expanding the RG $\beta$-functions to second order in the 
coupling constants, so that it is not allowed
to extrapolate this expression to large values of $g_0$.

To find out whether in the spinless two-chain system sufficiently strong
interchain-backscattering  
can give rise to a confinement transition where the renormalized effective
interchain hopping $t_{\bot}^{\ast} \propto k^+ - k^-$ vanishes, 
we use here a generalization of 
the collective field functional RG approach
with momentum transfer cutoff
developed in Ref.~[\onlinecite{Schuetz05}].
It turns out that with this approach we can 
analyze the regime where the dimensionless
interchain backscattering interaction $g_0$ is of the
order of unity. 
The crucial point is that from the weak coupling analysis \cite{Ledowski05}
we know that the confinement transition is 
driven by interchain backscattering,
so that it is natural to decouple  the interaction in this
scattering channel via a suitable bosonic Hubbard-Stratonovich field.
Simple truncations in the resulting mixed boson-fermion theory
correspond to infinite resummations
in an expansion of the RG $\beta$-functions
in powers of $g_0$.

\section{Effective low-energy model}

We consider  spinless fermions
confined to two chains that are coupled by weak interchain hopping $t_{\bot}$.
The kinetic energy of the two-chain system is diagonalized
by forming symmetric (bonding band) and antisymmetric (antibonding band)
combinations
of the eigenstates associated with isolated chains. 
Denoting by $\epsilon_k$  the energy dispersion of a 
single chain in the absence
of interchain hopping,
the energy dispersion of the non-interacting system
is $\epsilon_{ k}^{\sigma}  = \epsilon_k - \sigma t_{\bot}$,
where  $\sigma = + 1$ labels the bonding band and $\sigma =-1$ labels
the antibonding band.
It is useful to think of $\sigma$ as a pseudospin label \cite{Fabrizio93},
in which case $t_{\bot} =h $ corresponds to a uniform magnetic field $h$ 
in the $z$-direction. 

The problem of finding the low-energy
properties of two coupled metallic chains has been studied
previously by many 
authors \cite{Dusuel03,Ledowski05,Fabrizio93,Brazovskii85,Bourbonnais91,Kusmartsev92,Finkelstein93,Boies95,Balents96,Arrigoni98,Ledermann00,Louis01,Caron02,Bourbonnais04,Nickel06,Tsuchiizu06}.
However, the problem of self-consistently calculating the
true Fermi surface has only recently been addressed \cite{Dusuel03,Ledowski05,Louis01}.
At low energies all 
possible scattering processes in the spinless two-chain system
can be  divided into four different 
classes~\cite{Fabrizio93}: (1) forward
scattering processes, parameterized  in terms of 
three different  coupling constants $f^{++}$, $f^{--}$ and $f^{+-} = f^{-+}$, where the labels
indicate the band of the fermions involved in the scattering process;
(2) interband backward scattering, which in pseudospin language corresponds
to transverse spin-exchange, so that we shall call the corresponding
dimensionful coupling constant $J^{\bot}$ 
(the associated dimensionless coupling
$g_0$ will be introduced in Sec.~\ref{subsec:relevant});
(3)  pair-tunneling, which can also
be viewed as interband Umklapp 
scattering, parameterized in terms
of a coupling constant by $u$;
and finally (4)  intraband Umklapp scattering, 
which is expected to be important only at commensurate fillings.
Neglecting the latter process
and setting for simplicity
$f^{++} = f^{--}$, the low-energy interactions can
be expressed in terms of  four  marginal coupling constants
$f = \frac{1}{2}( f^{+-} + f^{++})$,
$J^{\parallel} =  \frac{1}{2}( f^{+-} - f^{++} )$,
$ J^{\bot}$, and $u$.
In the bonding-antibonding basis the
system can then be modeled by the following
effective Euclidean  action in pseudospin notation,
 \begin{eqnarray}
 S [ \bar{\psi} , \psi ] & = & \sum_{\sigma } \int_K
 ( - i \omega + \xi_{ k}^{ \sigma } )  \bar{\psi}_{ K }^{\sigma} 
 \psi_{ K}^{ \sigma}
 \nonumber
 \\
 &  & \hspace{-15mm} + \frac{1}{2} \int_{\bar{K}} 
 \left[ f (\bar{k} ) \bar{\rho}_{ \bar{K} } 
 \rho_{\bar{K}}
  -  J^{\parallel} (\bar{k} ) 
 \bar{m}_{ \bar{K}} m_{ \bar{K}} \right]
  \nonumber
 \\
  &   & \hspace{-15mm}  + \int_{ \bar{K}} 
 \left[ u ( \bar{k} )\left( 
 \bar{s}_{\bar{K}} \bar{s}_{-\bar{K}} +  s_{\bar{K}} s_{-\bar{K}} \right) 
 - 2 J^{\bot} ( \bar{k} ) \bar{s}_{ \bar{K} } s_{\bar{K}} \right]
 \; ,
 \nonumber
 \\
 & &
 \label{eq:action1}
 \end{eqnarray}
where $\xi_{ k}^{ \sigma} = \epsilon_k - \mu - \sigma h $, and
we have introduced the following composite fields,
 \begin{subequations}
 \begin{eqnarray}
 \rho_{\bar{K}} & = & \sum_\sigma \int_K \bar{\psi}_{ K}^{ \sigma}
 \psi_{ K + \bar{K}}^{ \sigma}
 \; ,
 \\
m_{\bar{K}} & = & \sum_\sigma   \sigma  \int_K  \bar{\psi}_{ K }^{\sigma}
 \psi_{ K + \bar{K} }^{ \sigma}
 \; ,
 \\
s_{\bar{K}} & = & \int_{{K}} \bar{\psi}_{ K }^{ -} \psi_{ K+ \bar{K} }^{+}
 \; .
 \end{eqnarray}
 \end{subequations}
We use the 
imaginary time formalism at zero temperature and
have introduced
collective labels $K =  ( k , i \omega )$
for fermionic fields and
$\bar{K} = ( \bar{k} , i \bar{\omega} )$ for bosonic fields, with the notation
$\int_K = \int \frac{dk d\omega}{(2\pi)^2}$. 
Note that the Fourier components of the density and the longitudinal
spin-density field satisfy
 $\rho_{-K} = \rho_K $ and  $m_{-K} = m_K$, while 
the spin-flip field $s_K$ 
is complex and do not have this symmetry.
The  interaction functions $f ( \bar{k} )$, 
$J^{\parallel} ( \bar{k} ) $, $J^{\bot} ( \bar{k} )$, and $u ( \bar{k} )$
should be considered 
as phenomenological low-energy couplings
which are only non-zero for
$ | \bar{k} | \leq \Lambda_0 \ll k_F$.
Hence, these couplings should not be directly compared with
the bare coupling constant in the Hubbard model \cite{footnotehubbard}.
The signs and normalizations in Eq.~(\ref{eq:action1}) are chosen such that
for $J^{\bot} = J^{\parallel}$ the model has
rotational invariance in pseudo-spin space, and
that for the Hubbard model all  couplings are
positive \cite{footnotehubbard}.
However, in our effective low energy model 
there is no reason to expect 
rotational invariance in pseudospin-space, so that in general
$J^{\parallel} ( \bar{k} ) \neq J^{\bot} ( \bar{k} )$.

\begin{figure}[tb]    
   \centering
% \psfrag{g}{$J^{\bot}$}
% \psfrag{f}{$J^{\parallel}$}
%  \vspace{7mm}
%     \includegraphics{fig1.eps}
      \epsfig{file=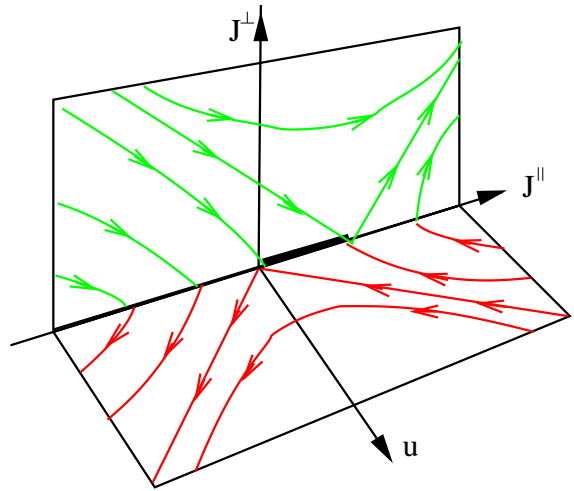,width=75mm}
%  \vspace{4mm}
  \caption{%
(Color Online) Qualitative behavior of the weak coupling RG flow
of the model (\ref{eq:action1}) in the space of coupling constants $J^{\parallel}$,
$J^{\bot}$ and $u$. The thick black line is the line of fixed points
describing the stable Luttinger liquid phase.
}
    \label{fig:flowLL}
  \end{figure}

The model defined in (\ref{eq:action1}) is still quite complicated and contains
many interaction processes which are not essential for the confinement transition.
In fact, from our previous weak 
coupling analysis~\cite{Ledowski05} we know that
the dominant renormalization of the difference between the Fermi points is due to
the interchain backscattering process described 
by the coupling $J^{\bot} ( \bar{k} )$.
In this work we study a minimal  model describing the confinement transition
 by simply neglecting the forward scattering
interactions $f ( \bar{k} )$ and $J^{\parallel} ( \bar{k} )$, as well as the
pair tunneling coupling $u ( \bar{k} )$ in Eq.~(\ref{eq:action1}). 
However it is known~\cite{Brazovskii85,Boies95} 
that sufficiently strong pair tunneling  can destabilize the Luttinger liquid phase; 
we shall come back to this point in Sec.~\ref{sec:conclusion}.
In pseudospin language, our model then describes a one-dimensional
spin $S=1/2$
Fermi system subject to a uniform magnetic field in $z$-direction with an attractive
ferromagnetic spin exchange involving only the transverse ($xy$) spin components.
The latter tends to align the spins in the direction perpendicular to the
magnetic field.  The Fermi surface renormalization is essentially
determined by the competition between the $xy$-exchange interaction and the
external constraint imposed by the
uniform magnetic field, which tends to align the spins along the
$z$-axis.
The phase diagram of the model (\ref{eq:action1}) in the space of all couplings
has been discussed by Fabrizio \cite{Fabrizio93}.
The qualitative behavior of the weak coupling RG flow
in the space of couplings spanned by  $J^{\parallel}$,
$J^{\bot}$ and $u$ is shown in Fig.~\ref{fig:flowLL}.
Obviously, there is a finite regime in coupling space where the 
spinless two-chain system is a stable Luttinger liquid, with gapless excitations.
In this work we shall assume that the qualitative fixed point structure 
suggested by the weak coupling analysis remains correct even in the strong
coupling regime. We can therefore choose the bare parameters such that the system
belongs to the basin of attraction of the Luttinger liquid fixed point manifold.

At low energies we may further simplify our model
(at least in the deconfined phase) by linearizing the energy
dispersion at the Fermi surface, which 
for our model consists of  four points $\alpha k^{  \sigma}$,
where the chirality index $\alpha = \pm 1$ labels the left/right 
Fermi point.
Note that the true Fermi points are defined via
 \begin{equation}
 \epsilon_{ \alpha {k}^{ \sigma} } - \mu - \sigma h  + \Sigma^{\sigma} 
 ( \alpha k^{ \sigma} , i0 ) =0
 \; ,
 \label{eq:FSdef}
 \end{equation}
where $\Sigma^{\sigma} (  \alpha k^{ \sigma} , i0 )$ 
is the exact self-energy for vanishing
frequency and for momenta at the true
Fermi surface $\alpha k^{\sigma}$ of the interacting system.
To linearize the energy dispersion at the true Fermi surface, we add 
and subtract from the
non-interacting  energy dispersion the counter-term
\begin{equation}
\mu^{\sigma}_0 = -  \Sigma^{\sigma} (  \alpha k^{ \sigma} , i0  ) \; ,
 \label{eq:counterdef}
\end{equation}
 and approximate
 \begin{eqnarray}
 \xi_{ \alpha k^{ \sigma} + k}^{ \sigma} & = & 
\epsilon_{ \alpha k^{ \sigma} + k } - \mu - \sigma h   
 \nonumber
 \\
&  & \hspace{-16mm} = 
\epsilon_{ \alpha k^{ \sigma} + k } - \mu - \sigma h   
+ \Sigma^{\sigma} (  \alpha k^{ \sigma} , i0 )
-
\Sigma^{\sigma} (  \alpha k^{ \sigma} , i0  )
 \nonumber
 \\
& & \hspace{-16mm} \approx
\alpha v^{\sigma}_0 k  + \mu^{\sigma}_0
 \; ,
 \end{eqnarray}
where $v_0^{\sigma}$ is the bare Fermi velocity at the
true Fermi surface for  pseudospin $\sigma$.
In analogy with the
definition of the couplings in the 
Tomonaga-Luttinger model~\cite{Solyom79}, we now generalize 
the interaction by introducing chirality
indices 
$ J^{\bot}  \rightarrow J^{\bot}_{\alpha \alpha^{\prime}}   $.
We shall refer to the diagonal processes $J^{\bot}_{ \alpha \alpha }$
as chiral interactions (these are called $g_4$ processes in the
Tomonaga-Luttinger model).
Similarly,  off-diagonal elements $J^{\bot}_{\alpha , -{\alpha}}$
will be  called non-chiral processes (corresponding to the $g_2$-processes 
in the Tomonaga-Luttinger model).
Defining new fields 
 \begin{equation}
  \psi^{\sigma}_{ K \alpha} = \psi_{ \alpha k^{ \sigma} + k , i \omega}^{ \sigma}
 \; ,
 \end{equation}
we replace the action (\ref{eq:action1}) 
by the following effective low-energy action describing the
physics of the 
confinement transition in our system of two spinless metallic chains,
 \begin{eqnarray}
 S [ \bar{\psi} , \psi ] & = & \sum_{\sigma, \alpha } \int_K
 ( - i \omega + \alpha v^{\sigma}_0 k  + \mu^{\sigma}_0 ) 
 \bar{\psi}^{\sigma}_{ K \alpha} \psi^{\sigma}_{ K \alpha}
 \nonumber
 \\
 & - &  2 \sum_{ \alpha \alpha^{\prime}}  
 \int_{\bar{K}} J^{\bot}_{ \alpha \alpha^{\prime} }  
\bar{s}_{ \bar{K} \alpha} s_{\bar{K} \alpha^{\prime}}
 \; ,
 \label{eq:action2}
 \end{eqnarray}
where it is understood that the $\bar{k}$-integration
has a momentum transfer cutoff $ |\bar{k} | \leq \Lambda_0 \ll k_F$, and
 \begin{eqnarray}
s_{ \bar{K} \alpha} & = & \int_K \bar{\psi}^{-}_{ K \alpha} 
 \psi^{+}_{ K+ \bar{K} \alpha}
 \; .
 \end{eqnarray}

\section{Exact RG flow equations}

\subsection{Hubbard Stratonovich transformation}

\begin{figure}[tb]    
   \centering
% \psfrag{t1}{$T >T_c$}
%  \vspace{7mm}
%     \includegraphics{fig1.eps}
      \epsfig{file=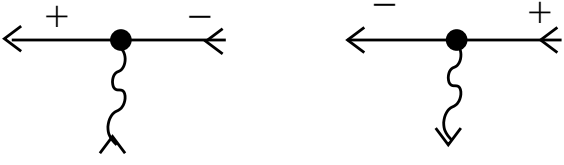,width=75mm}
%  \vspace{4mm}
  \caption{%
Bare interaction vertices of the action $S_1 [ \Phi ]$ given in Eq.~(\ref{eq:S1}).
The fermionic fields $\psi^{\sigma}$ and $\bar{\psi}^{\sigma}$ are denoted by
solid arrows, with the spin-projection $\sigma = \pm 1$ written next to the arrows. 
Bosonic spin-flip fields $\chi$ and $\bar{\chi}$ are denoted by
wavy arrows. Incoming arrows denote $\psi^{\sigma}$ and $\chi$, while
outgoing arrows correspond to the conjugate fields
$\bar{\psi}^{\sigma}$ and $\bar{\chi}$.
}
    \label{fig:vertexS1}
  \end{figure}

Because the confinement transition is a strong coupling phenomenon,
the usual weak coupling RG approach based on the expansion
in powers  of $J^{\bot}_{\alpha \alpha^{\prime}}$ is not suitable. 
To develop a RG approach which does not rely on a weak coupling expansion,
we decouple the spin-flip interaction with the help
of a complex Hubbard-Stratonovich field $\chi_{\alpha}$.
For convenience we  collect all fields in a super-field,
 \begin{equation}
 \Phi = ( \psi_{\alpha}^{+} , \bar{\psi}_{\alpha}^{+},
\psi_{\alpha}^{-} , \bar{\psi}_{\alpha}^{-}, \chi_{\alpha} , \bar{\chi}_{\alpha} )
 \; .
\end{equation}
Taking into account that there are two chiralities $\alpha = \pm 1$,
our super-field has totally
12 components (8 fermionic and 4 bosonic ones).
The ratio of the partition functions with and without interactions can then be written as
 \begin{equation}
 \frac{ \cal{Z}}{{\cal{Z}}_0 } = \frac{ \int{\cal{D}} [ \Phi ] e^{ - S_0 [ \Phi ] - S_1 [ \Phi ] } }{
 \int{\cal{D}} [ \Phi ] e^{ - S_0 [ \Phi ] }}
 \; ,
 \end{equation}
with the Gaussian part of the effective action given by
 \begin{eqnarray}
  S_0 [ \Phi ] & = &
\sum_{\sigma, \alpha } \int_K
 ( - i \omega + \alpha v^{\sigma}_0 k  + \mu^{\sigma}_0 ) 
 \bar{\psi}^{\sigma}_{ K \alpha} \psi^{\sigma}_{ K \alpha}
 \nonumber
 \\ 
  &+ & \frac{1}{2}  \sum_{ \alpha \alpha^{\prime} }  
\int_{\bar{K}} [ {\bf{J}}^{\bot} ]^{-1}_{\alpha 
 \alpha^{\prime}} \bar{\chi}_{\bar{K} \alpha} \chi_{\bar{K} \alpha^{\prime} }
\; .
 \label{eq:Gauss}
 \end{eqnarray}  
Here ${\bf{J}}^{\bot}$ is a  matrix
in chirality space with matrix elements given by
${J}^{\bot}_{\alpha \alpha^{\prime}}$, and the interaction is
\begin{eqnarray}
  S_1 [ \Phi ] & = &
\sum_{ \alpha } \int_{\bar{K}} 
\bigl[
 \bar{s}_{\bar{K} \alpha} \chi_{\bar{K} \alpha} +
 s_{\bar{K} \alpha}  \bar{\chi}_{\bar{K} \alpha}  \bigr]
 \; .
 \label{eq:S1}
 \end{eqnarray}
A graphical representation of the
bare interaction vertices in Eq.~(\ref{eq:S1})
is shown in Fig.~\ref{fig:vertexS1}.
The coupled RG flow equations for the one-line irreducible vertices
of the above mixed boson-fermion theory can be
obtained as a special case of the general flow equations given in
Ref.~[\onlinecite{Schuetz05}].

\subsection{Functional RG flow equations in momentum transfer cutoff scheme}

In order to calculate the true Fermi surface, we need to know
the exact counter-term $\mu^{\sigma}_0 
= - \Sigma ( \alpha k^{\sigma} , i 0  )$, which can be 
obtained from the flowing self-energy 
$\Sigma_{\Lambda}^{\sigma} ( K , \alpha )$
in the limit of vanishing infrared cutoff $\Lambda \rightarrow 0$.
The form of the flow equation 
for $\Sigma_{\Lambda}^{\sigma} ( K , \alpha )$
depends on the
RG method employed. Here we use the hierarchy of functional RG 
equations for the one-line irreducible vertices \cite{Wetterich93,Morris94}
of mixed boson-fermion models developed in Ref.~[\onlinecite{Schuetz05}].
A similar approach involving both fermionic and bosonic fields has been
developed in Refs.~[\onlinecite{Wetterich02,Baier04}].
In principle, one can also obtain the flowing self-energy
within the purely fermionic parameterization of the
hierarchy of flow equations \cite{Kopietz01,Ledowski05,Salmhofer01}.
However, with the usual truncations necessary
in  this approach it is not possible to reach the strong coupling regime.

\begin{figure}[tb]    
   \centering
% \psfrag{a}{$\sigma$}
% \psfrag{b}{$- \sigma$}
% \psfrag{d}{$\huge{\delta_{ \sigma ,-}}$}
%\psfrag{u}{$\delta_{ \sigma ,+}$}
%  \vspace{7mm}
      \epsfig{file=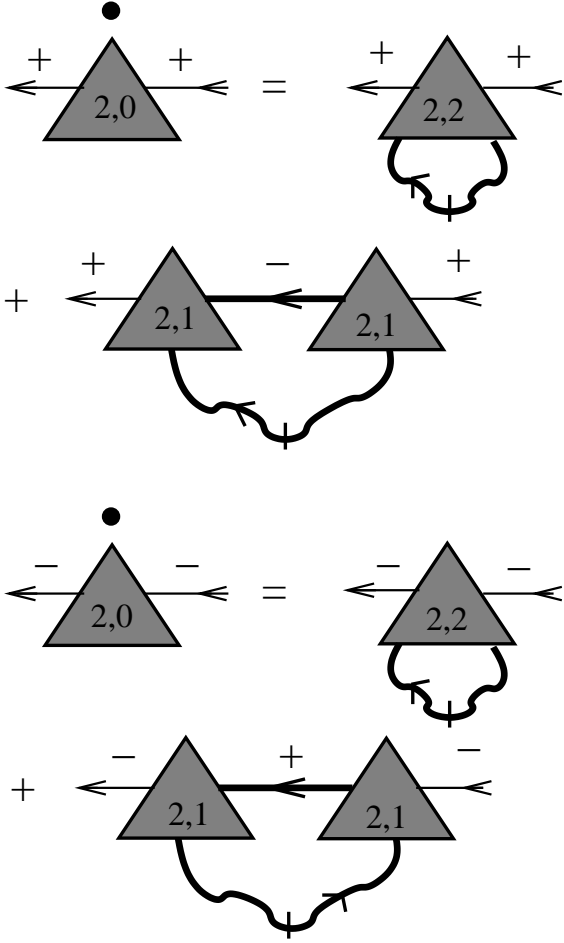,width=75mm}
%  \vspace{4mm}
  \caption{%
Exact flow equation for the fermionic self-energy $\Sigma^{\sigma}_{\Lambda} 
( K , \alpha)$
in the momentum transfer
cutoff scheme. 
The thick solid arrow is the flowing fermionic Green function and
the thick wavy line with a slash is the flowing single scale 
spin-flip propagator defined in Eq.~(\ref{eq:bossingle}).  
The  one-line irreducible vertices are represented by shaded triangles. 
A label $(n,m)$ inside a shaded triangle means that the vertex has
$n$ fermionic and $m$ bosonic external legs.
}
    \label{fig:flowsigma}
  \end{figure}

\begin{figure}[tb]    
   \centering
% \psfrag{t1}{$T >T_c$}
%  \vspace{7mm}
      \epsfig{file=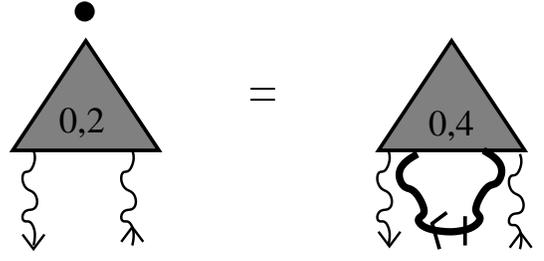,width=70mm}
%  \vspace{4mm}
  \caption{%
Exact flow equation for the spin-flip susceptibility  in the momentum transfer
cutoff scheme.
}
    \label{fig:flowPi}
  \end{figure}

\begin{figure}[tb]    
   \centering
% \psfrag{t1}{$T >T_c$}
%  \vspace{7mm}
      \epsfig{file=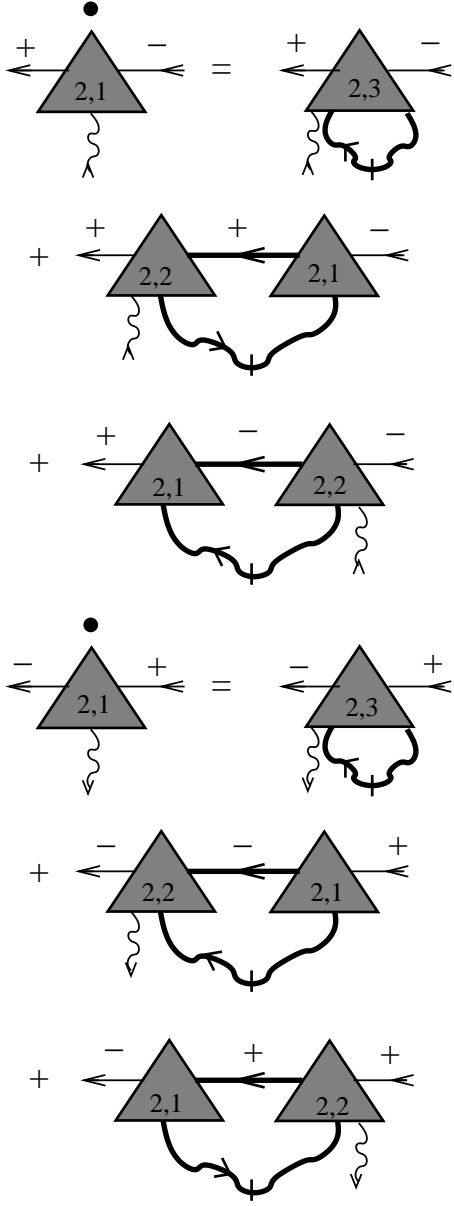,width=60mm}
%  \vspace{4mm}
  \caption{%
Exact flow equation for the  spin-flip vertices 
in the momentum transfer
cutoff scheme.
}
    \label{fig:flowvert}
  \end{figure}

In the momentum transfer cutoff scheme \cite{Schuetz05}
we impose a cutoff $\Lambda$ only on the momentum $ \bar{k} $ transfered by the
collective bosonic field. The resulting RG flow equation 
for the fermionic self-energy $\Sigma_{\Lambda}^{\sigma} ( K , \alpha )$ is
shown graphically in Fig.~\ref{fig:flowsigma}.
The corresponding analytic expression is
\begin{eqnarray}
\partial_{\Lambda} \Sigma^{\sigma}_{\Lambda} ( K, \alpha) &  &
 \nonumber
 \\
 & &  \hspace{-10mm} =
 \int_{ \bar{K} } \dot{F}^{ \sigma \bar{\sigma}}_{\Lambda} ( \bar{K} ,  \alpha  ) 
 \Gamma^{(2,2)}_{\Lambda} ( K  \sigma , -K  \sigma ; 
 \bar{K} , - \bar{K} , \alpha )
\nonumber
 \\
 &  & \hspace{-10mm} +
 \int_{ \bar{K} } \dot{F}^{ \sigma \bar{\sigma}}_{\Lambda} ( \bar{K} ,  \alpha  ) 
G^{ \bar{\sigma}}_{\Lambda} ( K + \bar{K} + \alpha \sigma \Delta,  \alpha )
 \nonumber
 \\
 &  &  \hspace{-5mm} \times 
 \Gamma^{(2,1) }_{\Lambda} ( K \sigma ; K + \bar{K}, \bar{\sigma} ; - \bar{K}, \alpha)
 \nonumber
 \\
 & & \hspace{-5mm} \times 
\Gamma^{(2,1)}_{ \Lambda} ( K + \bar{K}, \bar{\sigma} ;  K, \sigma ; \bar{K} , \alpha)
 \; .
 \label{eq:RGselfvertex}
 \end{eqnarray}
Here 
 $G_{\Lambda}^{\sigma} ( K , \alpha )$ is
the flowing fermionic single-particle Green function
for a given
pseudospin  $\sigma$ and chirality index $\alpha$.
We use the notation $\bar{\sigma} = - \sigma$ and measure
the wave-vectors $k$
with respect to the true Fermi momenta $\alpha k^{\sigma}$, defining
 \begin{equation}
 G_{\Lambda}^{\sigma} ( K , \alpha )
 = G_{\Lambda}^{\sigma} ( \alpha k^{\sigma} + k , i \omega )
 \; ,
 \end{equation}
and
 \begin{equation}
 G^{ \bar{\sigma}}_{\Lambda} ( K  + \alpha \sigma \Delta,  \alpha )
 =  G^{ \bar{\sigma}}_{\Lambda} (  \alpha k^{\bar{\sigma}} + k
  + \alpha \sigma \Delta,   i \omega  )
 \; . 
\end{equation}
The shift 
 $ \alpha \sigma \Delta = \alpha \sigma (k^{+} - k^- )$ in the argument
of  $G_{\Lambda}^{\bar{\sigma}} $ in Eq.~(\ref{eq:RGselfvertex})
is due to the fact that in  $G_{\Lambda}^{\bar{\sigma}} (K , \alpha )$ 
the wave-vector $k$ is measured with respect to a different reference point than
in $G_{\Lambda}^{{\sigma}} ( K , \alpha )$.
The function $\dot{F}^{ \sigma \bar{\sigma} }_{\Lambda} (  \bar{K} \alpha  )$
in Eq.~(\ref{eq:RGselfvertex}) is the
single scale bosonic spin-flip
propagator, which is defined by
 \begin{equation}
   \dot{F}^{ \sigma \bar{\sigma} }_{\Lambda} (  \bar{K} \alpha  )  = 
 - \delta ( | \bar{k} | - \Lambda )
    [ {\bf{F}}^{ \sigma \bar{\sigma}}_{\Lambda} ( \bar{K}  )]_{\alpha \alpha}
 \; ,
 \label{eq:bossingle}
 \end{equation}
where 
$ {\bf{F}}^{ \sigma \bar{\sigma}}_{\Lambda} ( \bar{K}  ) $ is a matrix
in chirality space whose inverse has the matrix elements
 \begin{equation}
  [ {\bf{F}}^{\sigma \bar{\sigma}}_{\Lambda} ( \bar{K} ) ]^{-1}_{\alpha \alpha^{\prime} }  = 
  [  2 {\bf{J}}^{\bot}  ]^{-1}_{\alpha \alpha^{\prime} } - 
 \delta_{\alpha \alpha^{\prime}} 
 \Pi_{ \Lambda}^{\sigma \bar{\sigma}} ( \bar{K} , \alpha  )
 \; ,
 \end{equation}
where $\Pi_{ \Lambda}^{\sigma \bar{\sigma}} ( \bar{K} , \alpha  )$ is the flowing spin-flip
susceptibility.
In the momentum transfer cutoff scheme, the RG flow of
$\Pi_{ \Lambda}^{\sigma \bar{\sigma}} ( \bar{K} , \alpha  )$
is driven by the one-line irreducible vertex with four external boson legs, as shown
in Fig.~\ref{fig:flowPi}. 
The vertices $ \Gamma^{(2,1) }_{\Lambda} 
( K \sigma ; K^{\prime} \bar{\sigma} ;  \bar{K} , \alpha)$ in 
Eq.~(\ref{eq:RGselfvertex}) are the flowing spin-flip vertices with two fermionic and
one bosonic external legs.  
In the momentum transfer cutoff scheme these vertices 
satisfy the exact flow equations  shown in graphically in Fig.~\ref{fig:flowvert}, with initial 
condition
 \begin{equation}
 \Gamma^{(2,1)}_{ \Lambda_0 } ( K \sigma ; 
 K^{\prime} \bar{\sigma} ; \bar{K}, \alpha )  =  1 
 \; .
 \label{eq:Gammainitial}
 \end{equation}
Finally, the vertex $\Gamma^{(2,2)}_{\Lambda}$
on the right-hand side of Eq.~(\ref{eq:RGselfvertex}) is the one-line irreducible
vertex with two fermionic and two bosonic external legs.
We do not give the flow equation for this vertex, because
purely bosonic vertices with more than two external legs
and  mixed vertices with two fermionic and more than one bosonic external leg
have negative scaling dimensions and are irrelevant~\cite{Schuetz05}.
We expect that their effect can be implicitly taken into account by re-defining
the numerical values of the relevant and marginal couplings \cite{Polchinski84}.
% Of course, the purely fermionic vertex 
% with four external legs as also marginal, but in
% the momentum transfer cutoff scheme this vertex 
% does not couple to the flow of the fermionic 
% self-energy.

The initial condition for 
the fermionic self-energy at scale $\Lambda = \Lambda_0$ is simply
 \begin{equation}
 \Sigma^{\sigma}_{\Lambda_0} ( K , \alpha )  =  0
 \; .
 \label{eq:sigmainitial}
 \end{equation}
Similarly, the vertices with two fermion legs and more than one
boson leg 
appearing on the right-hand side of the flow equation for the
spin-flip vertex shown in
Fig.~\ref{fig:flowvert} also vanish at the initial scale.
However,
the price we pay for introducing a cutoff only in the momentum transfer  
are non-trivial initial conditions  for the purely bosonic vertices, 
which are initially given by closed fermion loops \cite{Schuetz05}.
In particular,
the loop with two external boson legs
is initially given by the non-interacting spin-flip susceptibility,
 \begin{eqnarray} 
 \Pi^{\sigma \bar{\sigma}}_{ \Lambda_0} ( \bar{K} , \alpha  )  & = & 
 - \int_K
 G_{\Lambda_0}^{\sigma} ( K , \alpha ) G_{\Lambda_0}^{\bar{\sigma}} 
 ( K + \bar{K} + \alpha \sigma \Delta, \alpha )
\; ,
 \nonumber
 \\
 & &
 \label{eq:Piinitial}
 \end{eqnarray}
where for  our model with linear energy dispersion,  
 \begin{equation} 
G_{\Lambda_0}^{\sigma} ( K , \alpha ) = \frac{ 1 }{ i \omega - \alpha v^{\sigma}_0 k  
 - \mu^{\sigma}_0 }
 \; .
 \label{eq:G0lin}
 \end{equation}
Denoting by
\begin{equation}
 \Delta_{0} = 
k^{+}_{0} - k^{-}_{0}
 \label{eq:delta0def}
 \end{equation}
the distance between the Fermi momenta $k_0^{+}$ and $k_{0}^-$ in the absence
of interactions, the relation between the true distance 
 $\Delta = k^{+} - k^{-}$
and $\Delta_{0}$ can be expressed in terms of the counter-terms
$\mu^{\sigma}_0 =
- \Sigma ( \alpha k^{\sigma} , i0 , \alpha )$ 
as follows,
 \begin{equation}
 \Delta =
  \Delta_{0} 
  + \left[  \frac{ \mu^{+}_0}{ v^+_0}
 - \frac{ \mu^{- }_0 }{ v^-_0} \right]
 \; ,
 \label{eq:DeltaDelta0}
 \end{equation}
see also Eq.~(\ref{eq:FSdef2}) below.
Using   Eqs.~(\ref{eq:G0lin})  and (\ref{eq:DeltaDelta0}) 
we can explicitly evaluate Eq.~(\ref{eq:Piinitial}),
\begin{eqnarray} 
 \Pi^{\sigma \bar{\sigma}}_{ \Lambda_0} ( \bar{K} , \alpha  )  & = & 
  \frac{  1  }{ 2 \pi v^{\sigma}_0 }
 \frac{  v^{\bar{\sigma}}_0 ( \sigma \Delta_{0}  + \alpha  \bar{k})  }{ 
 v^{\bar{\sigma}}_0 ( \sigma \Delta_{0} + \alpha \bar{k} ) 
 - i \bar{\omega} }
\; .
 \label{eq:Piinitial2}
 \end{eqnarray}

\subsection{Rescaled flow equations and classification of vertices}

To classify the various vertices according to their relevance,
it is useful to make them dimensionless by multiplying them
with a suitable power of the running cutoff $\Lambda$.
Following Ref.~[\onlinecite{Schuetz05}],
we define dimensionless fermionic labels
$Q = ( q , i \epsilon ) = ( k/ \Lambda , i \omega / \Omega_{\Lambda} )$, 
and  bosonic ones
$\bar{Q} = ( \bar{q} , i \bar{\epsilon} ) = ( \bar{k} / \Lambda, i \bar{\omega} / \Omega_{ \Lambda} )$. Here 
$\Omega_{\Lambda} = v_F \Lambda$, where
$v_F $ is some average Fermi velocity.
For simplicity we shall write the above
relations as $Q  = K / \Lambda$ and $\bar{Q} = \bar{K} / \Lambda$.
We consider all rescaled quantities  as  functions of the
logarithmic flow parameter $l = \ln ( \Lambda_0 / \Lambda)$.

In order to define the Fermi surface within the framework
of the RG, we subtract the counter-term
$\Sigma^{\sigma}  ( \alpha k^{ \sigma} , i 0 , \alpha)  
= - \mu^{\sigma}_0$ from the flowing self-energy
and then rescale~\cite{Kopietz01},
 \begin{eqnarray}
 \tilde{\Sigma}_{l}^{ \sigma} ( Q , \alpha ) 
  & = & \frac{ Z^{\sigma}_{ l}}{ \Omega_{\Lambda} } 
 \left[ \Sigma_{\Lambda}^{ \sigma} 
 ( K , \alpha ) -  
\Sigma^{\sigma}  ( \alpha k^{ \sigma}, i 0 ) \right]
 \nonumber
 \\
 & = &
  \frac{ Z^{\sigma}_{ l}}{ \Omega_{ \Lambda} } 
 \left[ \Sigma_{\Lambda}^{ \sigma} 
 ( \Lambda Q  , \alpha )  + \mu^{\sigma}_0  \right]
 \; .
 \label{eq:sigmasub}
 \end{eqnarray}
Here $Z^{\sigma}_l$ is the flowing wave-function renormalization factor,
which is  defined in terms of the flowing self-energy as follows,
\begin{equation}
 Z^{\sigma }_{ l} = 1 + \left. \frac{ \partial 
\tilde{\Sigma}_l^{\sigma} ( 0 , i \epsilon , \alpha )}{
 \partial ( i \epsilon ) } \right|_{  \epsilon =0}
 \; .
 \end{equation}
The corresponding rescaled fermionic propagator is
 \begin{equation}
 \tilde{G}_l^{\sigma} ( Q, \alpha ) = \frac{ \Omega_{\Lambda}}{Z^{ \sigma }_{ l} }
 G^{ \sigma}_{\Lambda} ( \Lambda Q , \alpha )
 \; .
 \label{eq:Gscale}
 \end{equation} 
The rescaled self-energy satisfies
the exact  RG flow equation\cite{Kopietz01,Ledowski03}
 \begin{equation}
 \partial_l \tilde{\Sigma}_{l}^{ \sigma} ( Q , \alpha )     = 
( 1 - \eta^{\sigma }_{ l} - q \partial_q - \epsilon \partial_{\epsilon} )  
 \tilde{\Sigma}_{l}^{ \sigma} ( Q , \alpha ) 
 + \dot{{\Gamma}}_{ l }^{\sigma} ( Q , \alpha) 
\; ,
 \label{eq:flowsigma}
 \end{equation}
where the flowing anomalous dimension of the Fermi fields is
 \begin{equation}
 \eta_l^{\sigma} = - \partial_l \ln Z^{\sigma}_l =
 \left. - \frac{ \partial  \dot{{\Gamma}}^{\sigma}_{l } ( 0, i  \epsilon , \alpha)}{
 \partial ( i \epsilon) } \right|_{ \epsilon =0}
 \; ,
 \label{eq:etadef}
 \end{equation}
and  the  function $ \dot{{\Gamma}}_{ l }^{\sigma} ( Q , \alpha) $
follows from Eq.~(\ref{eq:RGselfvertex}),
 \begin{eqnarray}
 \dot{{\Gamma}}_{ l }^{\sigma} ( Q , \alpha)
 & = & \frac{ Z^{ \sigma }_{ l}}{\Omega_{\Lambda}}[ - \Lambda \partial_{\Lambda}
 \Sigma_{\Lambda}^{\sigma} ( K, \alpha ) ]
 \nonumber
 \\
 & = & 
 \int_{\bar{Q}} 
 \dot{\tilde{F}}^{ \sigma \bar{\sigma}}_l ( \bar{Q}  , \alpha  ) 
 \tilde{\Gamma}^{(2,2)}_l ( Q , \sigma ; -Q \sigma ; \bar{Q} , - \bar{Q} , \alpha )
 \nonumber
 \\
 &  + &\int_{\bar{Q}} 
 \dot{\tilde{F}}^{ \sigma \bar{\sigma}}_l ( \bar{Q}  , \alpha  ) 
\tilde{G}^{ \bar{\sigma}}_{l} ( Q + \bar{Q} + \alpha \sigma \tilde{\Delta}_l^{\ast},  \alpha )
 \nonumber
 \\
 &  &  \times 
 \tilde{\Gamma}^{(2,1) }_{ l} ( Q , \sigma ; Q + \bar{Q}, \bar{\sigma} ; - \bar{Q} , \alpha)
 \nonumber
 \\
 &  &  \times 
 \tilde{\Gamma}^{(2,1)}_{ l } ( Q + \bar{Q}, \bar{\sigma} ;  Q \sigma ; \bar{Q} , \alpha)
 \; ,
 \label{eq:dotgamma2}
 \end{eqnarray}
where
 \begin{equation}
 \tilde{\Delta}_l^{\ast} = \frac{ \Delta}{\Lambda} =  \frac{ \Delta}{\Lambda_0} e^l
 \label{eq:deltastardef} 
\end{equation}
is the rescaled true difference between the Fermi points.
The rescaled bosonic 
single scale propagator follows from Eq.~(\ref{eq:bossingle}),
 \begin{eqnarray}
 \dot{\tilde{F}}^{ \sigma \bar{\sigma}}_l ( \bar{Q} , \alpha  ) & = &
 - \frac{ \nu_0}{\bar{Z}_l } \Lambda \dot{F}^{\sigma \bar{\sigma}}_{\Lambda}
 ( \Lambda \bar{Q} , \alpha )
 \nonumber
 \\
 & = & \delta ( | \bar{q} | -1 )
    [ \tilde{{\bf{F}}}^{ \sigma \bar{\sigma}}_{l} ( \bar{Q}  )]_{\alpha \alpha}
 \; ,
 \label{eq:bossinglescale}
 \end{eqnarray}
with
 \begin{equation}
  [ \tilde{{\bf{F}}}^{\sigma \bar{\sigma}}_{l} ( \bar{Q} ) ]^{-1}_{\alpha \alpha^{\prime} }  = 
   \bar{Z}_l [  2 \nu_0 {\bf{J}}^{\bot}  ]^{-1}_{\alpha \alpha^{\prime} } - 
 \delta_{\alpha \alpha^{\prime}} 
 \tilde{\Pi}_{ l }^{\sigma \bar{\sigma}} ( \bar{Q} , \alpha  )
 \; ,
 \label{eq:tildeFpropdef}
 \end{equation}
and the rescaled spin-flip susceptibility
 \begin{equation}
 \tilde{\Pi}_l^{ \sigma \bar{\sigma}} ( \bar{Q}, \alpha ) =
 \frac{ \bar{Z}_l}{ \nu_0 } \Pi_{\Lambda}^{\sigma \bar{\sigma}} ( \Lambda \bar{Q} , \alpha )
 \; .
  \label{eq:Pisfrescale}
 \end{equation}
Here $\bar{Z}_l$ is the 
wave-function renormalization factor associated with the
bosonic spin-flip field $\chi_{\alpha}$, and the constant
$\nu_0$ with units of inverse velocity has been introduced  
to make all  rescaled vertices dimensionless.
The rescaled spin-flip vertex in Eq.~(\ref{eq:dotgamma2}) is
 \begin{eqnarray}
 \tilde{\Gamma}^{(2,1)}_{ l } ( Q, \sigma ;  
 Q^{\prime}, \bar{\sigma}; \bar{Q} , \alpha ) & = & 
 \left[ {  \bar{Z}_l   }/{ \nu_0  }
 \right]^{1/2} \left[ { \Lambda Z_{l}^{+} Z_{l}^- }/{\Omega_{\Lambda} }
\right]^{1/2}
 \nonumber
 \\
 &     & \hspace{-15mm} \times
{\Gamma}^{(2,1)}_{ \Lambda } ( \Lambda Q , \sigma ;  
 \Lambda Q^{\prime}, \bar{\sigma} ; \Lambda \bar{Q}, \alpha )
 \; ,
 \label{eq:vertexrescale}
 \end{eqnarray} 
and the rescaled vertex with two external fermion and two boson legs is
 \begin{eqnarray}
\tilde{\Gamma}^{(2,2) }_l ( Q^{\prime}, \sigma ; Q , \sigma ; \bar{Q}^{\prime} ,  
\bar{Q} ,  \alpha )
 & = & \Lambda  Z_l^{\sigma}   ( \bar{Z}_l / \nu_0 ) 
 \nonumber
 \\
 &    & \hspace{-38mm}  \times
{\Gamma}^{(2,2) }_{ \Lambda } ( \Lambda Q^{\prime} , \sigma;  
 \Lambda Q, \sigma  ; \Lambda \bar{Q}^{\prime}, \Lambda \bar{Q}, \alpha )
 \; .
 \label{eq:vertexrescale22}
 \end{eqnarray} 
For convenience we now choose
$\nu_0 = \Lambda / \Omega_{\Lambda} = 1/v_F$ so that the
prefactor on the right-hand side 
of Eq.~(\ref{eq:vertexrescale}) reduces to 
$[ \bar{Z}_l Z^{+}_l Z^-_l ]^{1/2}$.

Let us now classify the various vertices according to their relevance.
First of all, the key quantity to
obtain the counter-terms $\mu^{\sigma}_0$ is
the momentum- and  frequency independent part of the rescaled self-energy
$ \tilde{\Sigma}_{l}^{ \sigma} ( Q , \alpha ) $ defined in Eq.~(\ref{eq:sigmasub}),
which we call
 \begin{equation}
 r^{\sigma }_{ l} =   \tilde{\Sigma}_{l}^{ \sigma} ( 0 ,  \alpha ) =
 \frac{ Z^{\sigma}_{ l}}{ \Omega_\Lambda } 
 \left[ \Sigma_{\Lambda}^{ \sigma} 
 ( 0 , \alpha )  + \mu^{\sigma}_0  \right]
 \; .
 \end{equation}
The couplings $r_l^{\sigma}$ satisfy the exact flow equation
\begin{equation}
 \partial_l r_{ l }^{ \sigma} = ( 1 - \eta_{l}^{ \sigma} )  r_{ l }^{ \sigma}
 + \dot{{\Gamma}}^{\sigma}_{l } ( 0, \alpha) 
\; ,
 \label{eq:rlflow}
 \end{equation}
with initial condition 
 \begin{equation}
r_0^{\sigma} = \frac{\mu^{\sigma}_0 }{ \Omega_{\Lambda_0}}
 =  - \frac{ \Sigma^{\sigma} ( \alpha k^{\sigma} , i0  )}{ v_F \Lambda_0 }
 \; .
 \label{eq:initialr0}
\end{equation}
There are two marginal couplings related to the self-energy. The first is  the
wave-function renormalization factor $Z_l^{\sigma}$, which according to
Eq.~(\ref{eq:etadef}) is related to the flowing anomalous dimension via
 \begin{equation}
 \partial_l Z^{\sigma}_{ l} = - \eta^{\sigma }_{ l}  Z^{\sigma }_{ l} 
 \; .
 \label{eq:Zflow}
 \end{equation}
The second is
the dimensionless Fermi velocity renormalization factor
 \begin{equation}
 \tilde{v}_l^{\sigma} = Z_l^{\sigma} +  \left. \frac{ \partial 
 \tilde{\Sigma}_l^{\sigma} ( q , i 0 , \alpha )}{
 \partial  ( \alpha q) } \right|_{  q =0}
 \; ,
 \end{equation}
 which satisfies the exact flow equation
 \begin{equation}
 \partial_l \tilde{v}_l^{\sigma}  =  -  \eta_l^{\sigma}  \tilde{v}_l^{\sigma} +
 \left.  \frac{ \partial \dot{\Gamma}^{\sigma}_l ( q , i 0, \alpha) 
 }{\partial (\alpha q) } 
 \right|_{ q=0} 
\label{eq:vlflow}
 \; .
 \end{equation}
If we retain only relevant and marginal couplings, the
rescaled fermionic propagator with energy dispersion linearized at the
true Fermi surface is given by
\begin{equation}
 \tilde{G}_l^{\sigma} ( q , i \epsilon,  \alpha ) \approx \frac{1}{ 
 i \epsilon - \alpha \tilde{v}_l^{\sigma} q - r_l^{\sigma}}
 \; .
 \label{eq:Gscalerel1}
 \end{equation} 
Apart from $Z_l^{\sigma}$ and $\tilde{v}_l^{\sigma}$, 
the third marginal coupling of our model is
the momentum- and frequency-independent part
of the rescaled spin-flip vertex defined in Eq.~(\ref{eq:vertexrescale}),
 \begin{equation}
 \gamma_l = \tilde{\Gamma}^{(2,1)}_{l} ( 0, \sigma;0, \bar{\sigma} ;0 , \alpha)
 \; .
 \end{equation}
It satisfies a flow equation of the form
 \begin{equation}
 \partial_l \gamma_l = - \frac{ \bar{\eta}_l + \eta_l^+ + \eta_l^- }{2}
 \gamma_l + \dot{\Gamma}^{(2,1)}_l
 \; , 
 \label{eq:gammabotflow}
 \end{equation}
where  $\bar{\eta}_l = - \partial_l \ln \bar{Z}_l$ is the flowing anomalous 
dimension of the spin-flip field, and
the inhomogeneity  $\dot{\Gamma}^{(2,1)}_l$ depends on the
irrelevant higher interaction vertices involving more
than one external boson leg shown  in Fig.~\ref{fig:flowvert}.
In particular, 
from the right-hand side of Eq.~(\ref{eq:vertexrescale22})
it is clear that the vertex $\tilde{\Gamma}_l^{(2,2) }$ 
with two external fermion  and two boson legs
is irrelevant with scaling dimension $-1$. 
This and the higher order irrelevant vertices vanish at the initial scale $\Lambda_0$ 
and we shall set them equal to zero, expecting that their effect
can be implicitly taken into account
via a redefinition of the numerical values of the 
relevant and marginal couplings \cite{Polchinski84}.
An exception is the vertex $\Gamma^{(0,4)}_{\Lambda}$ 
involving four external bosonic legs, which
according to Fig.~\ref{fig:flowPi} 
drives  the flow of the spin-flip susceptibility
 in the momentum transfer cutoff scheme.
In contrast to the other irrelevant vertices, the vertex
$\Gamma^{(0,4)}_{\Lambda}$ is finite at the initial 
scale $\Lambda = \Lambda_0$, where it
reduces to a symmetrized closed fermion loop \cite{Schuetz05}.
Below we shall propose a simple approximate procedure to take
the renormalization of the spin-flip susceptibility generated by this vertex
into account.
Finally, we note that 
the vertex with four external fermionic legs is also marginal, but
in the momentum transfer cutoff scheme it does not  directly couple
to the flow of the fermionic self-energy.

\subsection{Defining the Fermi surface within the functional RG}
\label{subsec:FS}

The general method to obtain the counter-terms necessary to construct the
true Fermi surface within the framework of the
functional RG has been 
developed in Refs.~[\onlinecite{Kopietz01,Ledowski03,Ledowski05}].
Let us briefly recall the main idea.
As long as the  flowing anomalous dimensions
$\eta_l^{\sigma}$ of the Fermi fields remains smaller than unity  
for $l \rightarrow \infty$, we may
define the true Fermi surface self-consistently from the
requirement that the relevant couplings $r^{ \sigma }_{ l}$ associated with the
fermionic self-energy approach finite limits
for $l \rightarrow \infty$.
This requires fine tuning of the
initial values $r^{\sigma}_0$, which defines a relation between
$r^{\sigma}_0$ and the flowing couplings
on the entire RG trajectory. 
In higher dimensions, where the Fermi surface is a continuum, infinitely
many relevant couplings $r_l ( {\bd{k}}_F )$ have to be fine tuned
to define the Fermi surface. In the usual classification of critical fixed points, the
Fermi surface thus corresponds to a multicritical point of infinite order.
Once the proper  initial values $r_0^{\sigma}$ are known,
the exact self-energy $\Sigma^{\sigma} ( \alpha k^{\sigma} , i0)$ can
be constructed using Eq.~(\ref{eq:initialr0}),
 \begin{equation}
 \Sigma^{\sigma} ( \alpha k^{\sigma} , i 0 ) = - \mu^{\sigma}_0 = - v_F \Lambda_0 r_0^{\sigma}
 \; .
 \end{equation}      
The requirement that $r_l^{\sigma}$ flows into a RG fixed point
implies for the initial values\cite{Ledowski03},
 \begin{equation}
  r_{0}^{ \sigma} = - \int_{0}^{\infty} dl e^{ - ( 1 - 
\bar{\eta}^{\sigma}_l ) l }
 \dot{{\Gamma}}^{\sigma}_{l } (0)
 \; ,
 \label{eq:selfcon}
 \end{equation}
where
 \begin{equation}
 \bar{\eta}_l^{\sigma} = \frac{1}{l} \int_0^{l} d t \eta_{t}^{\sigma}
 \end{equation}
is the average of the flowing anomalous dimension along the RG trajectory,
and we have written  $\dot{{\Gamma}}^{\sigma}_{l } (0) =
\dot{{\Gamma}}^{\sigma}_{l } (0, i0 , \alpha)$ to emphasize
that this quantity is actually independent of  the chirality index $\alpha$.
For our effective  model with linear energy dispersion we obtain
for the Fermi point distance at constant chemical potential
[see also Eq.~(\ref{eq:FSdef})],
 \begin{eqnarray}
 \tilde{\Delta} & = & \tilde{\Delta}_0
  +  
 \left[ \frac{r_0^+}{\tilde{v}_0^{+}} - \frac{ r_0^-}{\tilde{v}_0^-}  \right]
 \nonumber
 \\
 &  & \hspace{-10mm} = \tilde{\Delta}_0
 - \sum_{\sigma} \frac{\sigma}{ \tilde{v}_0^{\sigma} }
 \int_{0}^{\infty} dl e^{ - ( 1 - 
\bar{\eta}^{\sigma}_l ) l }
 \dot{{\Gamma}}^{\sigma}_{l } (0, \alpha)
 \; ,
 \label{eq:FSdef2}
 \end{eqnarray}
where we have defined
 \begin{equation}
 \tilde{\Delta} = \frac{ k^{+} - k^{-}}{\Lambda_0} \; , \;
\tilde{\Delta}_0 = \frac{ k^{+}_0 - k^{-}_0}{\Lambda_0}
 \; .
 \end{equation}

\section{Calculation of the true Fermi surface}
\label{sec:Calculation}

\subsection{Truncation based on relevance} 

Because a possible confinement transition is expected to be a strong-coupling 
phenomenon,
the usual perturbative weak coupling RG \cite{Ledowski05} is not
sufficient.
We  therefore propose an alternative truncation scheme
based on the truncation according
to relevance in the RG sense. 
In our model we have to keep track of the RG flow of the
two relevant couplings $r_l^{\sigma}$, $\sigma = \pm 1$, and the
marginal couplings $Z_l^{\sigma}$, $\tilde{v}_l^{\sigma}$, and $\gamma_l$.
The couplings $r_l^{\sigma}$, $Z_l^{\sigma}$ and
$\tilde{v}_l^{\sigma}$ associated with the fermionic Green function
satisfy the flow equations
given in Eqs.~(\ref{eq:rlflow}, \ref{eq:Zflow}, \ref{eq:vlflow}).
The function 
 $ \dot{{\Gamma}}_{ l }^{\sigma} ( Q , \alpha) $ appearing on the
right-hand side of these equations
is in general given in Eq.~(\ref{eq:dotgamma2});  approximating
the fermionic Green function by  Eq.~(\ref{eq:Gscalerel1}) and
the spin-flip vertex by its momentum- and frequency independent part
$\gamma_l$, we obtain
 \begin{eqnarray}
 \dot{{\Gamma}}_{ l }^{\sigma} (q , i \epsilon ,   \alpha)
 & = &   \int \frac{ d \bar{q}  d \bar{\epsilon} }{( 2 \pi )^2} \delta ( | \bar{q} | -1 )
 \nonumber
 \\ 
 & & \hspace{-23mm} \times \frac{ \gamma_l^2
 [ {\bf{\tilde{F}}}_l^{\sigma \bar{\sigma}} ( \bar{q}, i \bar{\epsilon} ) ]_{ \alpha \alpha }
   e^{ i \bar{\epsilon} 0^+} }{ 
 i (  \bar{\epsilon} + \epsilon ) - \alpha \tilde{v}_l^{\bar{\sigma}} ( 
 \bar{q} +q )  - \sigma \tilde{v}^{\bar{\sigma}}_l \tilde{\Delta}^{\ast}_l 
 -r^{\bar{\sigma}}_l 
}
 \; .
 \label{eq:dotgamma3}
 \end{eqnarray}
Here $\tilde{\Delta}_{l}^{\ast}$
is the rescaled true difference between the Fermi 
points defined in  Eq.~(\ref{eq:deltastardef}), and
the rescaled bosonic spin-flip propagator 
 ${\bf{\tilde{F}}}_l^{\sigma \bar{\sigma}} ( \bar{Q} )$
is defined in Eq.~(\ref{eq:tildeFpropdef}).

To calculate the Fermi surface, we need
additional flow equations for the
marginal part of the spin-flip vertex $\gamma_l$ and for the
flowing
spin-flip susceptibility   
$\tilde{\Pi}^{ \sigma \bar{\sigma}}_l  ( \bar{Q}, \alpha ) $.
As far as $\gamma_l$ is concerned,
we note from Fig.~\ref{fig:flowvert} that the inhomogeneity
$\dot{\Gamma}^{(2,1)}_l$ 
in Eq.~ (\ref{eq:gammabotflow}) which drives the flow of
$\gamma_l$
involves vertices with two fermionic and more than one  bosonic external legs.
These vertices are
irrelevant and  vanish at the initial scale $\Lambda_0$, so that it is
reasonable to neglect them. We therefore set $\dot{\Gamma}^{(2,1)}_l = 0$.
We shall also neglect
the bosonic wave-function renormalization, setting 
$\bar{Z}_l =1$. In this approximation the flow of the rescaled spin-flip vertex
is driven by the fermionic wave-function renormalization,
 \begin{equation}
\partial_l \gamma_l  =  - \frac{\eta_l^+ + \eta_l^-}{2} \gamma_l
\label{eq:gammalflow}
 \; .
 \end{equation}

Before discussing the spin-flip susceptibility
$\tilde{\Pi}^{ \sigma \bar{\sigma}}_l  ( \bar{Q}, \alpha ) $, note that
Eqs.~(\ref{eq:vlflow}) and (\ref{eq:dotgamma3}) imply for the Fermi velocity 
renormalization factor
 \begin{equation}
 \partial_l \tilde{v}_l^{\sigma}  = - \eta_l^{\sigma} ( \tilde{v}_l^{\sigma} 
 -  \tilde{v}_l^{\bar{\sigma}} )
 \; ,
 \end{equation}
which yields for the difference
 \begin{equation}
 \partial_l ( \tilde{v}_l^+ - \tilde{v}_l^-) =
 - ( {\eta}_l^+ + \eta_l^- )  ( \tilde{v}_l^+ - \tilde{v}_l^-)
 \; .
 \label{eq:vdifflow}
 \end{equation}
Keeping in mind that
$\eta_l^{\sigma} \geq 0$,
Eq.~(\ref{eq:vdifflow}) implies that 
a small initial difference between the Fermi velocities
decreases under renormalization.
Thus, if the initial difference $v_0^+ - v_0^-$ is small
and negligible, it becomes even smaller as we iterate the RG.
Since the flow of the other couplings
is not sensitive to a small difference in the Fermi velocities,
it is consistent to approximate $v_0^{\sigma} \approx v_F$, so that
from now on we shall set $\tilde{v}_l^{\sigma} = 1$.

To close our system of flow equations, we need an equation
for the
rescaled  spin-flip susceptibility
$\tilde{\Pi}_l^{ \sigma \bar{\sigma}} ( \bar{Q} , \alpha )$, 
which in turn determines the flow of the spin-flip propagator as given
in Eq.~(\ref{eq:Pisfrescale}).
In the momentum transfer cutoff scheme, the flow 
of $\tilde{\Pi}_l^{ \sigma \bar{\sigma}} ( \bar{Q} , \alpha )$ 
is driven by the one-line irreducible vertex $\Gamma^{(0,4)}_{\Lambda}$
with four external bosonic legs, as shown in Fig.~\ref{fig:flowPi}.
Although this vertex is irrelevant, it is finite at the initial scale
$\Lambda_0$, in contrast to the higher order vertices
that drive the flow of the spin-flip vertex $\gamma_l$ shown
in Fig.~\ref{fig:flowvert}.
It is therefore important to take the renormalizations
of $\tilde{\Pi}_l^{ \sigma \bar{\sigma}} ( \bar{Q} , \alpha )$ 
due to $\Gamma^{(0,4)}_{\Lambda}$ at least approximately into account.
Guided by the initial condition  (\ref{eq:Piinitial2})
for the spin-flip susceptibility, we 
propose the following {\it{adiabatic approximation}},
 \begin{eqnarray}
 \tilde{\Pi}_l^{ \sigma \bar{\sigma}} ( \bar{Q} , \alpha ) & \approx &
  \frac{ \gamma_l^2 }{ 2 \pi  }
 \frac{   \sigma \tilde{\Delta}_{l}  + \alpha  \bar{q}  }{ 
   \sigma \tilde{\Delta}_{l} + \alpha \bar{q}  
 - i \bar{\epsilon} }
\; ,
 \label{eq:Pisfapprox}
 \end{eqnarray}
where
 \begin{equation}
 \tilde{\Delta}_{l} =  \tilde{\Delta}_l^{\ast} - ( r_l^{+} - r_l^- )
 \; .
 \label{eq:tildeDeltaldef}
 \end{equation}
Note that
Eq.~(\ref{eq:Pisfapprox}) preserves the
initial form of the spin-flip susceptibility
given in Eq.~(\ref{eq:Piinitial2}), but with the
initial gap $\tilde{\Delta}_0 = ( k_0^{+} - k_0^-)/\Lambda_0$  
is replaced by the flowing gap $\tilde{\Delta}_l$ at scale $l$, 
and an overall reduction of the amplitude
by the vertex correction $\gamma_l^2$. 
Indeed,  using Eq.~(\ref{eq:FSdef2}) we find
$\tilde{\Delta}_{ l=0} = \tilde{\Delta}_0 = ( k_0^{+} - k_0^-)/\Lambda_0$,
so that
for $l=0$
we recover from Eq.~(\ref{eq:Pisfapprox}) the rescaled version
of the initial condition (\ref{eq:Piinitial2}).
On the other hand,
using the fact that the $r_l^{\sigma}$ are fine tuned
to reach a finite limit for $l \rightarrow \infty$, we see that
$\Delta_l \rightarrow \Delta^{\ast}_l = e^l (k^+ - k^- ) / \Lambda_0 = e^l \tilde{\Delta}$
for $l \rightarrow \infty$.
% We may therefore define the flowing Fermi point difference
% $k_l^{+} - k_l^-$ via
%  \begin{equation}
%  \frac{ k_l^{+} - k_l^-}{\Lambda} = \tilde{\Delta}_l =
%  \frac{ k^{+} - k^-}{\Lambda} -( r^+_l - r^-_l )
%  \; .
%  \end{equation}
A justification for the adiabatic
approximation
(\ref{eq:Pisfapprox})  is given in the Appendix.

To simplify the integrals, it is convenient 
to slightly modify the denominator in the expression for
$\dot{{\Gamma}}_{ l }^{\sigma} (q , i \epsilon ,   \alpha)$
in Eq.~(\ref{eq:dotgamma3}),
 \begin{equation}
 \sigma \tilde{\Delta}_l^{\ast} + r_l^{\bar{\sigma}} =
 \sigma \tilde{\Delta}_l +  r_l^{{\sigma}} 
 \approx  \sigma \tilde{\Delta}_l 
 \; .
 \label{eq:rneglect}
 \end{equation}
We have checked numerically 
from the solution of the resulting equations that
this approximation is self-consistent
by verifying the neglected term $r_l^{\sigma}$ is indeed small.
We thus arrive at the following  approximation for the 
inhomogeneity $ \dot{{\Gamma}}_{ l }^{\sigma} (q , i \epsilon ,   \alpha)$
that controls the flow of the fermionic self-energy,
\begin{eqnarray}
 \dot{{\Gamma}}_{ l }^{\sigma} (q , i \epsilon ,   \alpha)
 & = &   \int \frac{ d \bar{q}  d \bar{\epsilon} }{( 2 \pi )^2} \delta ( | \bar{q} | -1 )
 \nonumber
 \\ 
 & & \hspace{-23mm} \times \frac{  \gamma_l^2
 [ {\bf{\tilde{F}}}_l^{\sigma \bar{\sigma}} ( \bar{q}, i \bar{\epsilon} ) ]_{ \alpha \alpha }
   e^{ i \bar{\epsilon} 0^+} }{ 
 i (  \bar{\epsilon} + \epsilon ) - \alpha  ( 
 \bar{q} +q )  - \sigma  \tilde{\Delta}_l 
}
 \; .
 \label{eq:dotgamma4}
 \end{eqnarray}
To be consistent the approximation (\ref{eq:rneglect})
we should also neglect the
flowing anomalous dimension $\eta_l^{\sigma}$ in
the flow equation (\ref{eq:rlflow}) for $r_l^{\sigma}$, because
an expansion of Eq.~(\ref{eq:dotgamma3}) 
in powers of  $r_l^{\sigma}$ leads to a cancellation
of the term $\eta_l^{\sigma} r_l^{\sigma}$.
The flow equation for $r_l^{\sigma}$ then reduces to
 \begin{equation}
 \partial_l r_{ l }^{\sigma}  =   r_{ l }^{\sigma} 
 + \dot{{\Gamma}}_{l }^{\sigma} (0) 
\; ,
 \label{eq:rflow4}
\end{equation}
where 
 \begin{eqnarray}
\dot{{\Gamma}}_{l }^{\sigma} ( 0) 
 & = &   \int \frac{ d \bar{q}  d \bar{\epsilon} }{( 2 \pi )^2} \delta ( | \bar{q} | -1 )
 \nonumber
 \\ 
 &  \times & \frac{  \gamma_l^2
 [ {\bf{\tilde{F}}}_l^{\sigma \bar{\sigma}} ( \bar{q}, i \bar{\epsilon} ) ]_{ \alpha \alpha }
 e^{ i \bar{\epsilon} 0^+} }{ 
 i   \bar{\epsilon}  - \alpha  
 \bar{q}   - \sigma  \tilde{\Delta}_l  
}
 \; .
 \label{eq:dotgamma04}
 \end{eqnarray}
Our general self-consistency equation (\ref{eq:FSdef2})
for the true Fermi point distance 
can then we written as an integral involving
the flow of the couplings on the entire RG trajectory, 
 \begin{eqnarray}
\tilde{\Delta} & = & \tilde{\Delta}_0
 -  \int_{0}^{\infty} dl e^{-l}
 \sum_{\sigma} \sigma
 \dot{{\Gamma}}^{\sigma}_{l } (0)
 \; .
 \label{eq:FSapprox1}
 \end{eqnarray}
Anticipating that within our approximations  $\eta_l^{\sigma}$ is
 independent of $\sigma$,  we may write
$\eta^{\sigma}_l = \eta_l$.  
The flow equation (\ref{eq:gammalflow})
for the spin-flip vertex  then reduces to
 \begin{equation}
 \partial_l \gamma_l = - \eta_l  \gamma_l
 \; ,
 \label{eq:gammabotflow2}
 \end{equation}
where from Eqs.~(\ref{eq:etadef}) and (\ref{eq:dotgamma4}) 
we find
 \begin{equation}
 \eta_l  = 
\int \frac{ d \bar{q}  d \bar{\epsilon} }{( 2 \pi )^2} \delta ( | \bar{q} | -1 ) \frac{  \gamma_l^2 
  [ {\bf{\tilde{F}}}_l^{\sigma \bar{\sigma}} ( \bar{q}, i \bar{\epsilon} ) ]_{ \alpha \alpha }
    }{ [
 i   \bar{\epsilon}  - \alpha   \bar{q}   - \sigma \tilde{\Delta}_l 
]^2 }
 \; .
 \label{eq:etaapprox2}
 \end{equation}
We thus arrive at 
a closed system of flow equations
for the two relevant couplings $r_l^{+}$ and $r_l^{-}$ and the two marginal
couplings $g_{n,l}$ and $g_{c,l}$.
We emphasize that our truncation does not rely
on a weak coupling expansion, which enables us to
study a possible confinement transition.

To give an explicit expression for the
bosonic spin-flip propagator, 
we define dimensionless bare couplings
(keeping in mind that we have chosen $\nu_0 = 1/v_F$),
 \begin{equation}
 2 \nu_0 J^{\bot}_{ \alpha \alpha} = 2 \pi g_{c,0}
 \; , \;
 2 \nu_0 J^{\bot}_{ \alpha, - \alpha} = 2 \pi g_{n,0}
 \; ,
 \end{equation}
and the  flowing couplings
 \begin{equation}
 g_{c,l}  =   \gamma_l^2 g_{c,0}
 \; , \; 
g_{n,l}  =   \gamma_l^2 g_{n,0}
 \label{eq:gnl}
 \; ,
 \end{equation}
which according to Eq.(\ref{eq:gammabotflow2})  satisfy the flow equations
 \begin{equation}
 \partial_l g_{c,l} = - 2 \eta_l g_{c,l }
 \; , \; 
 \partial_l g_{n,l} = - 2 \eta_l g_{n,l }
 \; .
 \label{eq:gflow}
 \end{equation}
The rescaled spin-flip propagator can then be written as
 \begin{widetext} 
\begin{eqnarray}
  \gamma_l^2 
 [ {\bf{\tilde{F}}}_l^{\sigma \bar{\sigma}} ( \bar{q}, i \bar{\epsilon} ) ]_{ \alpha \alpha }
 & = &
 \frac{  \gamma_l^2   2 \pi    [ g_{c,0}  -  
(g_{c,0}^2 - g_{n,0}^2) 2 \pi
  \tilde{\Pi}_l^{\sigma \bar{\sigma}} ( \bar{Q} , -\alpha )]}{1 -   g_{c,0}
 2 \pi [  \tilde{\Pi}_l^{\sigma \bar{\sigma}} ( \bar{Q} ,+  )
 +   \tilde{\Pi}_l^{\sigma \bar{\sigma}} ( \bar{Q} , -  ) ]
 +  (g_{c,0}^2 - g_{n,0}^2)   
 (2 \pi )^2 \tilde{\Pi}_l^{\sigma \bar{\sigma}} ( \bar{Q} , +  )
   \tilde{\Pi}_l^{\sigma \bar{\sigma}} ( \bar{Q} , -  )
}
 \nonumber
% \\
%  & = &  2 \pi
% \frac{  g_{c,l} 
% [ ( i \bar{\epsilon} - \sigma \tilde{\Delta}_l )^2 - \bar{q}^2 ]
% + (g_{c,l}^2 - g_{n,l}^2) [  \tilde{\Delta}_l^2 - \bar{q}^2 - i \bar{\epsilon}
% ( \sigma \tilde{\Delta}_l - \alpha \bar{q} )] }{
% [ i \bar{\epsilon} - \omega^+_l ( \bar{q}, \sigma \tilde{\Delta}_l ) ]
% [ i \bar{\epsilon} - \omega^-_l ( \bar{q} ,  \sigma \tilde{\Delta}_l ) ]
%  }
% \nonumber
 \\ & = &
 2 \pi
 ( i \bar{\epsilon} - \alpha \bar{q} - \sigma \tilde{\Delta}_l )
 \frac{ g_{c,l}
   ( i \bar{\epsilon} + \alpha \bar{q} - \sigma \tilde{\Delta}_l )
 + (g_{c,l}^2 - g_{n,l}^2) (  \alpha \bar{q} - \sigma \tilde{\Delta}_l ) 
 }{
 [ i \bar{\epsilon} - \omega^+_l ( \bar{q}, \sigma \tilde{\Delta}_l ) ]
 [ i \bar{\epsilon} - \omega^-_l ( \bar{q} ,  \sigma \tilde{\Delta}_l ) ]
  }
 \; ,
 \end{eqnarray}
where
 \begin{eqnarray}
 \omega^{\pm}_l  ( \bar{q} , x ) & = & x ( 1 - g_{c,l} ) \pm
 \sqrt{ x^2 g_{n,l}^2 + \bar{q}^2 [ ( 1 - g_{c,l} )^2 - g_{n,l}^2 ] }
 \nonumber
 \\
 & = &
x ( 1 - g_{c,l} ) \pm
 \sqrt{ x^2  ( 1 - g_{c,l} )^2 + ( \bar{q}^2 - x^2) [ ( 1 - g_{c,l})^2 - 
 g_{n,l}^2 ] }
 \; .
 \label{eq:omegapmdef}
 \end{eqnarray}
Noting that $\omega^{\pm}_l ( 0,x ) =  x ( 1 - g_{c,l} \pm g_{n,l} )$,
we see that
for small interaction strength 
both modes $\omega^+_l$ and $\omega^-_l$ are gapped.
However, the gap of the mode $\omega^{-}_l ( 0,x )$ vanishes for
$ g_{c,l} + g_{n,l} =1$, signaling a possible quantum phase transition
to a confined state.
In the present work we do not attempt to extend the RG beyond this
point, but focus
on the regime  $ g_{c,l} + g_{n,l} \leq 1$ were both modes are gapped.
The integrals in Eqs.~(\ref{eq:dotgamma04}) and (\ref{eq:etaapprox2})
can be  carried out analytically using the residue theorem, with the result
 \begin{eqnarray}
 \dot{\Gamma}_l^{\sigma} ( 0 ) & = & 
  2 \delta_{ \sigma ,-1} \Theta(  \tilde{\Delta}_l -1 ) g_{c,l}
 + \Theta (1- \tilde{\Delta}_l) 
% \nonumber
% \\
%& & \hspace{-13mm}
%\times
 \left[ g_{c,l} + \frac{ \sigma   \tilde{\Delta}_l   g_{n,l}^2}{ \sqrt{ 
 (1-g_{c,l})^2 - g_{n,l}^2 (1 - \tilde{\Delta}_l^2 ) } }
 \right] 
 \; ,
 \end{eqnarray}
and
 \begin{eqnarray}
 \eta_l & = &
 \frac{ \Theta ( 1 - \tilde{\Delta}_l )   
 g_{n,l}^2}{ \sqrt{ (1-g_{c,l})^2 - g_{n,l}^2 (1 - \tilde{\Delta}_l^2 )    }
 \left[ 1- g_{c,l} +  \sqrt{  (1-g_{c,l})^2 - g_{n,l}^2 (1 - \tilde{\Delta}_l^2 )   } \right]}
 \; .
 \label{eq:etares1}
 \end{eqnarray}
\end{widetext}

\subsection{Self-consistent one-loop approximation}
\label{subsec:relevant}

The above system of coupled equations can only be solved numerically.
However, if we neglect the flow of the coupling constants  on the 
right-hand sides of these equations, we can obtain an approximate analytical solution, 
which is useful to get a rough idea about the mechanism responsible
for the confinement transition. In this subsection we therefore 
set
 \begin{eqnarray}
  g_{ i, l} & \approx & g_{i ,0}
 \; ,
 \label{eq:gapprox}
 \\
  \tilde{\Delta}_l & \approx & \tilde{\Delta}_l^{\ast} \equiv 
  \tilde{\Delta} e^l 
 \; .
 \label{eq:Deltaapprox}
 \end{eqnarray}
We expect that these approximations over-estimate
the tendency towards confinement, because we
know from Eq.~(\ref{eq:gflow}) that the flowing couplings $g_{c,l}$ and $g_{n,l}$
are smaller than the bare ones.
The second approximation (\ref{eq:Deltaapprox})
is justified provided the trajectory integral 
(\ref{eq:FSapprox1}) is dominated by $l \gtrsim 1$ 
where $\tilde{\Delta}_l^{\ast}
 \gg | r_l^{\sigma} |$. 
Substituting $x  =
 \tilde{\Delta} e^{l}$ on the right-hand side of
Eq.~(\ref{eq:FSapprox1}) we obtain
 \begin{eqnarray}
\tilde{\Delta} 
 & = & \tilde{\Delta}_0 - \tilde{\Delta} 
\int_{\tilde{\Delta}}^{\infty} dx I ( x )
 \; ,
 \label{eq:FSapprox2}
 \end{eqnarray}
with
 \begin{eqnarray}
  I ( x ) & =&  -  \Theta ( x-1 )    \frac{2 g_{c,0}}{x^2}
 \nonumber
 \\
 &  &  \hspace{-20mm} + \Theta ( 1-x ) \frac{ 2 g_{n,0}^2}{x  
\sqrt{  
   (1-g_{c,0})^2 - g_{n,0}^2  (1-  x^2)  }
 }
 \; .
 \end{eqnarray}
The $x$-integration is elementary and 
we finally obtain
 \begin{equation}
 \tilde{\Delta}  = \frac{  \tilde{\Delta}_0  }{ 1 + R ( \tilde{\Delta}  ) }
 \; ,
 \label{eq:Rdef}
 \end{equation}
with
  \begin{eqnarray}
 R ( \tilde{\Delta}  ) & = & 
\int_{ {\tilde{\Delta}} }^{\infty} dx I ( x )
= - 2 g_{c,0}  
 \nonumber
 \\
 &   & \hspace{-15mm} + \frac{ 2 g_{n,0}^2}{ \sqrt{ (1  - g_{c,0}) ^2 - g_{n,0}^2} }
 \ln \left[ \frac{1 + \sqrt{ 1 + \frac{  \tilde{\Delta}^2  g_{n,0}^2}{ (1 - g_{c,0})^2 -g_{n,0}^2}}}{
 \tilde{\Delta} 
 \left( 1 + \sqrt{  \frac{ ( 1 - g_{c,0})^2}{ (1 - g_{c,0})^2 - g_{n,0}^2}} \right) } \right]
 \; .
 \nonumber
 \\
 & &
 \label{eq:Rres1}
\end{eqnarray}
This expression diverges for $g_{c,0} + g_{n,0}  \rightarrow  1$, 
corresponding 
to a confinement transition with $\tilde{\Delta} \rightarrow 0$.
Expanding 
$R ( \tilde{\Delta}  )$ to second order in the couplings we obtain
\begin{equation}
   R ( \tilde{\Delta} )  = 
- 2 g_{c,0} + 2 g_{n,0}^2 \ln (1 / \tilde{\Delta} ) + O (g_{i,0}^3 )
 \; .
 \label{eq:Rexpand}
 \end{equation}
Keeping in mind that in Ref.~[\onlinecite{Ledowski05}] 
we have neglected the chiral couplings and that here we have retained only
interchain backscattering, 
Eq.~(\ref{eq:Rexpand}) is consistent with
our previous weak-coupling result given in
Eq.~(\ref{eq:deltaprevious}).
From Eq.~(\ref{eq:Rexpand}) it is clear that
there is a competition between the chiral part $g_{c,0}$ and the
non-chiral part $g_{n,0}$ of the interaction.
The chiral part $g_{c,0}$
leads to a repulsion of the Fermi points, while the non-chiral part
$g_{n,0}$ generates an attraction and eventually triggers the confinement
transition, because
for sufficiently small $\tilde{\Delta}$ the logarithm
overwhelms the term linear in $g_{c,0}$.

\begin{figure}[tb]    
   \centering
% \psfrag{t1}{$T >T_c$}
%  \vspace{7mm}
%   \includegraphics{fig5.eps}    
    \epsfig{file=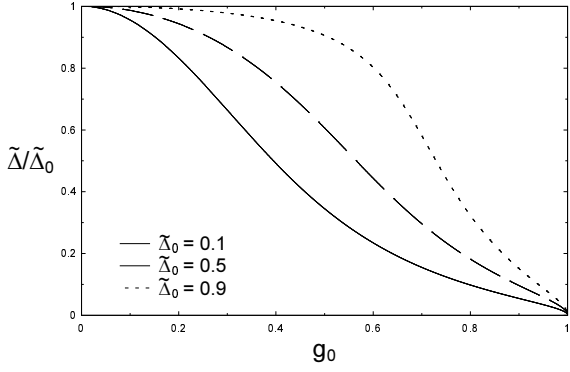,width=75mm}
%  \vspace{4mm}
  \caption{%
Numerical solution  of Eq.~(\ref{eq:Rdef}) for $g_{c,0} =0$ as function of
$g_0 = g_{n,0}$ for different values of $\tilde{\Delta}_0 = ( k^{+}_0 - k^{-}_0 )/\Lambda_0$.
}
    \label{fig:deltachiral}
  \end{figure}

To simplify the following analysis we shall from  now on restrict ourselves
to the special case $g_{c,0} =0$. This is sufficient
to study the confinement transition, which is driven by the
non-chiral part of the interaction.
Setting $g_0 = g_{ n , 0}$ the confinement 
transition occurs within the approximations in this section
at $g_0 =1$.
A numerical solution of Eqs.~(\ref{eq:Rdef}) and (\ref{eq:Rres1}) for the true
 $\tilde{\Delta}$ as function of $g_0$ is shown
in Fig.~\ref{fig:deltachiral}.
The confinement transition for $g_0 \rightarrow 1$ is clearly visible.
In fact, the behavior of $\tilde{\Delta}$ for $g_0 \rightarrow 1$ can 
be obtained analytically.
In this case
 $\tilde{\Delta} \ll \sqrt{ 1 - g_0^2}$, so that we may approximate
 \begin{eqnarray}
  R ( \tilde{\Delta}  ) & \approx &  
 \frac{2}{  \sqrt{ 1 -  g_0^2 } }
 \ln \left[ \frac{2  \sqrt{ 1 - g_0^2} }{  \tilde{\Delta}  }\right]
 \; .
 \label{eq:Rdiv}
 \end{eqnarray}
The self-consistency condition  (\ref{eq:Rdef}) for $\tilde{\Delta}$ 
then reduces  to
 \begin{equation}
 \tilde{\Delta}  \approx 
  \sqrt{ 1 - g_0^2}
    \left[ \frac{ \tilde{\Delta}_0}{ 2 \ln 
 \left( \frac{ 2 \sqrt{1 - g_0^2} }{  {\tilde{\Delta}} } \right) } \right] 
 \; .
 \label{eq:confinement}
 \end{equation} 
For  $\tilde{\Delta} \ll \sqrt{ 1 - g_0^2}$ the second factor in the square braces
of Eq.~(\ref{eq:confinement}) is small compared with unity, so that
it is consistent to take the limit $g_0 \rightarrow 1$ in this expression.
If we identify  $\Delta$ with the order parameter of the confinement transition
transition (with $\Delta \neq 0$ corresponding to the deconfined
phase), then Eq.~(\ref{eq:confinement}) predicts mean-field behavior with
logarithmic corrections.

Our simple one-loop approximation thus predicts
that for $g_0 \rightarrow 1$
there is a confinement transition where the true
Fermi point distance $\Delta$ collapses, corresponding to vanishing effective
interchain hoping $t_{\bot}^{\ast} =0$.
In pseudo-spin language,
the quantum critical point  $g_0=1$ corresponds
to a vanishing magnetization in $z$-direction, in spite of the
fact that there is a uniform magnetic field.
For $g_0 > 1$ our one-loop approximation
suggests that there is   long-range ferromagnetic order in 
$xy$-direction. However, in one dimension we do not expect  true long-range order,
so that fluctuations beyond the one-loop approximation should  be important.
It is therefore important to go beyond this approximation, which we shall do
in the following subsection.

\subsection{Including  the renormalization of the 
effective interaction}
\label{subsec:including}

\begin{figure}[tb]    
   \centering
% \psfrag{t1}{$T >T_c$}
%  \vspace{7mm}
%\includegraphics{fig6.eps}
     \epsfig{file=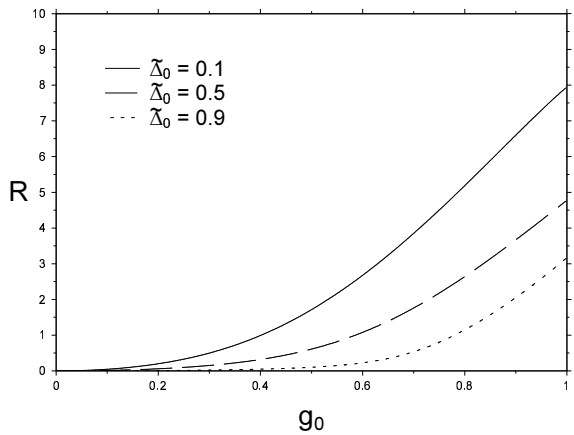,width=75mm}
%  \vspace{4mm}
  \caption{%
Numerical solution  of the renormalization factor $R ( \tilde{\Delta} )$
defined in  Eq.~(\ref{eq:Rres2})
as a function of the bare coupling $g_0$ for
different values of $\tilde{\Delta}_0$.
}
    \label{fig:Rres2}
  \end{figure}

\begin{figure}[tb]    
   \centering
% \psfrag{t1}{$T >T_c$}
%  \vspace{7mm}
      \epsfig{file=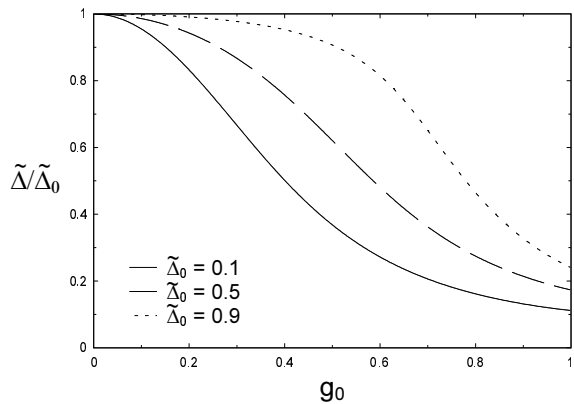,width=75mm}
%  \vspace{4mm}
  \caption{%
Numerical solution  of the true Fermi point distance $\tilde{\Delta} 
= ( k^{+} - k^- )/\Lambda_0$
as a function of the bare coupling $g_0$ for
different values of the bare distance $\tilde{\Delta}_0 = ( k^{+}_0 - k^-_0 )/\Lambda_0 $.
}
    \label{fig:deltatrueg0}
  \end{figure}

We now improve the above calculation by taking the flow of the
effective interaction into account.  
For simplicity, we focus again on the special case without
chiral interactions, so that we need to keep track only
of the flowing non-chiral interaction $g_{ n , l} = g_l$.
Furthermore, in the absence of chiral couplings $r_l^{\sigma} = \sigma r_l$.
The self-consistency equation (\ref{eq:FSapprox1})   
for the Fermi point distance then reduces to
 \begin{equation}
 \tilde{\Delta} = \tilde{\Delta}_0 - 
 \int_0^{\infty} dl e^{-l} 
 \frac{ 2 \Theta ( 1 - \tilde{\Delta}_l ) \tilde{\Delta}_l g_l^2 }{ \sqrt{ 1 - g_l^2 
 ( 1 - \tilde{\Delta}_l )^2 } }
 \; ,
 \label{eq:DeltaDelta}
 \end{equation}
where
 \begin{equation}
 \tilde{\Delta}_l = \tilde{\Delta}_l^{\ast} - 2 r_l  =
\tilde{\Delta} e^l - 2 r_l
 \label{eq:tildedeltalr}
 \; ,
 \end{equation}
and the flow of $r_l$ and $g_l$ is determined by
 \begin{eqnarray}
 \partial_l r_l  & = & r_l + A (  g_l,  \tilde{\Delta}_l )
 \; ,
 \label{eq:Arl}
 \\
 \partial_l g_l  & = & B ( g_l , \tilde{\Delta}_l )
 \; ,
 \label{eq:Brl}
 \end{eqnarray} 
with
 \begin{equation}
 A ( g_l , \tilde{\Delta}_l ) =
   \frac{   \Theta (1- \tilde{\Delta}_l)    \tilde{\Delta}_l   g_{l}^2}{ \sqrt{ 
 1  - g_{l}^2 (1 - \tilde{\Delta}_l^2 ) } }
 \; ,
 \label{eq:Adef}
 \end{equation}
and
 \begin{eqnarray}
 B ( g_l , \tilde{\Delta}_l ) & = & - 2 \eta_l g_l
 \nonumber
 \\
& & \hspace{-25mm} 
= 
 \frac{ - 2 \Theta ( 1 - \tilde{\Delta}_l )   
 g_{l}^3}{ \sqrt{ 1 - g_{l}^2 (1 - \tilde{\Delta}_l^2 )    }
 \left[ 1 +  \sqrt{  1 - g_{l}^2 (1 - \tilde{\Delta}_l^2 )   } \right]}
 \; .
 \label{eq:Bdef}
 \end{eqnarray}
The initial value $r_0$ has to be fine tuned such that
$\lim_{ l \rightarrow \infty} r_l $ remains finite. This leads to the
self-consistency equation (\ref{eq:DeltaDelta}) for the true Fermi point distance.
We emphasize again that 
our approximation scheme is not based on a weak coupling expansion,
so that the $\beta$-function given in
Eq.~(\ref{eq:Bdef}) is non-perturbative in the coupling $g_l$.
Instead of Eq.~(\ref{eq:Rres1}) we now obtain for the dimensionless renormalization 
factor $R ( \tilde{\Delta} )$ defined in Eq.~(\ref{eq:Rdef}),
 \begin{eqnarray}
 R ( \tilde{\Delta}  ) & = & \frac{2}{\tilde{\Delta}} \int_0^{\infty} d l e^{ - l } 
 A ( g_l , \tilde{\Delta}_l )
 \nonumber
 \\
 &  &  \hspace{-17mm} = 2
 \int_0^{\infty} d l 
  ( 1  -  2 e^{-l} r_l / \tilde{\Delta} )
 \frac{       \Theta ( 1 - \tilde{\Delta}_l )  g_l^2 }{
 \sqrt{ 1 -   g_l^2 ( 1 - \tilde{\Delta}_l^2 )} }
\; .
 \label{eq:Rres2}
 \end{eqnarray}
Note that formally the
confinement transition 
manifests itself via a divergence of
the function  $R (\tilde{\Delta} )$ 
for $\tilde{\Delta} \rightarrow 0$.
However, as shown in Fig.~\ref{fig:Rres2},
the renormalization factor remains now finite so that
also the self-consistent $\tilde{\Delta}$ 
is finite for $g_0 =1$, as shown in Fig.~\ref{fig:deltatrueg0}.
We conclude that
the confinement transition obtained in
the previous subsection is an artefact of
the approximations (\ref{eq:gapprox}) and (\ref{eq:Deltaapprox}).

\begin{figure}[tb]    
   \centering
% \psfrag{t1}{$T >T_c$}
%  \vspace{7mm}
      \epsfig{file=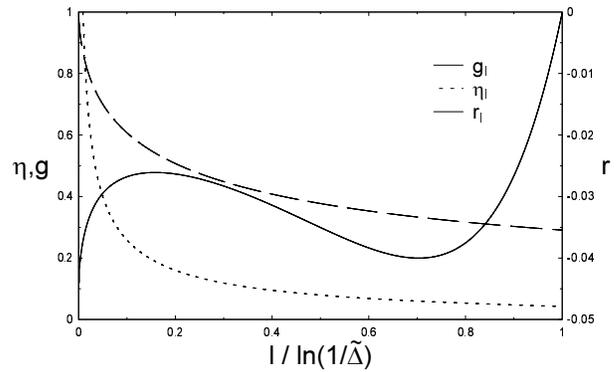,width=80mm}
%  \vspace{4mm}
  \caption{%
RG flow of the couplings $g_l$, $r_l$ and the anomalous dimension $\eta_l$ 
 as a function of the logarithmic flow parameter $l$ for
$\tilde{\Delta}_0 =0.1$ and  $g_0 =1$. 
}
    \label{fig:RGflow}
  \end{figure}

\begin{figure}[tb]    
   \centering
% \psfrag{t1}{$T >T_c$}
%  \vspace{7mm}
      \epsfig{file=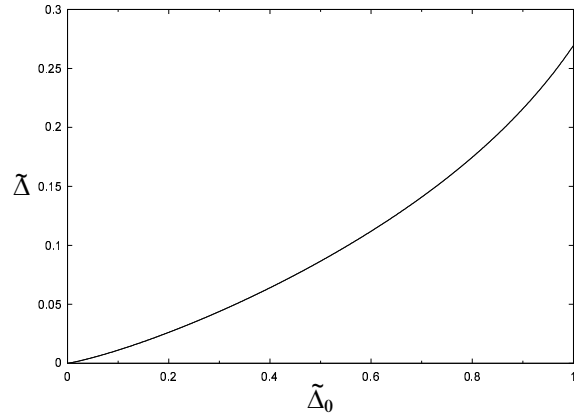,width=75mm}
%  \vspace{4mm}
  \caption{%
Numerical solution  of the true Fermi point distance  $\tilde{\Delta}$
for $g_0=1$ as a function of the bare distance $\tilde{\Delta}_0$.
}
    \label{fig:deltatrue_g1}
  \end{figure}

Let us take a closer look at  the point $g_0 =1$ where the one-loop 
approximation
predicts a confinement transition.
The RG flow of the couplings $g_l$ and $r_l$ as well as the flowing 
anomalous dimension $\eta_l$ for this case is shown in Fig.~\ref{fig:RGflow}.
One clearly sees that the initially large flowing anomalous dimension
drives the running coupling $g_l$  towards smaller values; 
eventually $g_l$ approaches  a finite limit for large $l$. Furthermore, the running 
coupling $r_l$
approaches its asymptotic value for large $l$ non-monotonously.
The true Fermi point distance $\Delta$ as a function of the
bare one for $g_0 =1$ is shown in Fig.~\ref{fig:deltatrue_g1}.
We see that  large interchain backscattering
strongly reduces the Fermi point distance, although the Fermi surface never collapses,
in agreement with  scenario suggested by the weak coupling analysis \cite{Ledowski05}.

\section{Conclusions}
\label{sec:conclusion}

In this work we have used a functional RG approach
to calculate self-consistently the true distance
$\Delta = k^{+} - k^{-}$ between the Fermi points
of the bonding and the antibonding band in
a system consisting of two chains of spinless fermions
connected by weak interchain hopping $t_{\bot}$. 
Using the insight from our earlier weak coupling analysis \cite{Ledowski05}
that the renormalization of the Fermi surface is essentially
determined by interchain backscattering, 
we have treated this scattering
process  non-perturbatively by representing it in terms of
a collective bosonic field $\chi$.
In pseudospin language, 
where $t_{\bot} = h$ corresponds to a uniform magnetic
field in $z$-direction and
the interchain backscattering interaction
corresponds to a ferromagnetic $xy$-interaction,
the field $\chi$ can be viewed as  a fluctuating transverse magnetic field, 
which competes with  the uniform field $h$ in $z$-direction.
A self-consistent  one-loop approximation 
predicts that for sufficiently strong interchain backscattering
there is indeed a quantum critical point were
the renormalized Fermi point distance $\Delta \propto t_{\bot}^{\ast} $ vanishes.
However, a more accurate calculation taking vertex corrections
and wave-function renormalizations into account 
shows that the renormalized $\Delta \propto t_{\bot}^{\ast}$
remains  finite.
This is in agreement with the expectation that a ferromagnetic $xy$-interaction
in a one-dimensional itinerant electron gas cannot
give rise to long-range ferromagnetic order.
Previous studies of the spinless two-chain problem \cite{Bourbonnais91,Caron02}
came to the conclusion that the system exhibits a confinement transition
if the anomalous dimension $\eta_0$ of the Luttinger liquid  for $t_{\bot}=0$
is unity. The important difference between these earlier works and our calculations is that
we have completely neglected pair tunneling. In a subsequent article~\cite{Ledowski06} 
we shall show how 
the inclusion of this process stabilizes again the flow of the interchain backward 
scattering and enhances the tendency towards confinement.
In the same article we shall
also show that our approach can be
generalized to study the
more interesting and physically more relevant 
confinement problem in an infinite array of
weakly coupled  metallic chains. In this case
the Fermi surface consists of two disconnected sheets, which
self-consistently develop completely flat sectors 
at the confinement transition. We have preliminary evidence
that in this case there exists a confined phase
where the renormalized interchain hopping vanishes.
The essential scattering process driving this transition is
the non-chiral part of the density-density interaction which transfers momentum
within a given  sheet of  the Fermi surface.

Finally we point out  that  this work
describes also some technical progress: 
we have been  able to 
find a sensible extrapolation of the weak coupling
functional RG approach to the strong coupling regime.
Our truncation strategy of the formally exact hierarchy
of functional RG equations relies on the classification
of the vertices according to their relevance. 
An obvious disadvantage of this approach is  that
we cannot give reliable error estimates, which
is a common feature of most truncations of the
coupled functional RG flow equations for the vertex functions. 
Note, however, that a similar truncation of the
functional RG equations for the interacting Bose gas
gave quite accurate results for the shift in the critical temperature\cite{Ledowski04}.
In the present
problem we know a priori  from the weak coupling analysis that
the physics is dominated by
a single scattering channel, the inter-chain backscattering.
However, an extension of our approach  
to  problems where several scattering 
channels compete seems to be possible.

\section*{ACKNOWLEDGMENTS}
We thank  Florian Sch\"{u}tz for sharing his insights
on the subtleties of the collective field functional
RG approach with us.

\appendix
\renewcommand{\theequation}{A.\arabic{equation}}
\renewcommand{\thesubsection}{A.\arabic{subsection}}

\section*{Appendix: Justification of the adiabatic approximation}

\begin{figure}[tb]    
   \centering
% \psfrag{t1}{$T >T_c$}
%  \vspace{7mm}
      \epsfig{file=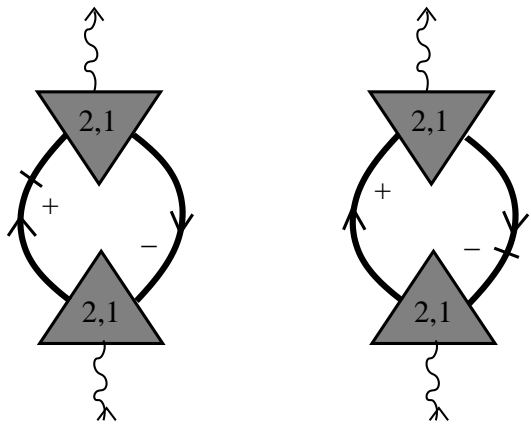,width=70mm}
%  \vspace{4mm}
  \caption{%
Additional diagrams contributing to the flow of the spin-flip susceptibility
in a cutoff scheme with a fermionic band-width cutoff. 
The solid arrow  with slash is the fermionic single-scale propagator.
}
    \label{fig:flowsuscepfermi}
  \end{figure}

We give here a  justification for the adiabatic approximation 
for the rescaled spin-flip
susceptibility given in Eq.~(\ref{eq:Pisfapprox}).
% The perhaps simplest way to arrive at this expression 
% is based on the 
% Dyson-Schwinger 
% equation~\cite{Schuetz05,ZinnJustin02} 
% for the spin-flip susceptibility, which is shown diagrammatically
% in Fig.~\ref{fig:DS}.
% %
% %
% \begin{figure}[tb]    
%    \centering
% % \psfrag{t1}{$T >T_c$}
% %  \vspace{7mm}
%       \epsfig{file=fig10.eps,width=60mm}
% %  \vspace{4mm}
%   \caption{%
% Dyson-Schwinger equation for the spin-flip susceptibility.
% }
%     \label{fig:DS}
%   \end{figure}
% %
% %
% Replacing in the Dyson-Schwinger equation
% the spin-flip vertex by unity,
% approximating the rescaled Green functions by Eq.~(\ref{eq:Gscalerel1}),
% and using the fact that within our approximations
% $Z_l^{+} Z_l^{-} = \gamma_l^2$,
% we arrive at Eq.~(\ref{eq:Pisfapprox}).
Let us therefore use a more general two-cutoff procedure
where we impose a
band-width cutoff $\Lambda^F_{\tau} = \Lambda_{0}^F e^{- \tau }$
on the fermionic propagator in addition to
the bosonic momentum transfer cutoff $ \Lambda_l = \Lambda_0 e^{- l}$.
All vertices and coupling constants then depend on both logarithmic
flow parameters $l $ and $\tau$.
Instead of Eq.~(\ref{eq:Gscalerel1}) 
the rescaled fermionic propagator can then be approximated by
\begin{equation}
 \tilde{G}_{l, \tau }^{\sigma} ( q , i \epsilon , \alpha ) \approx 
\frac{ \Theta ( 1 < | q| < e^{\tau} )}{ 
 i \epsilon - \alpha \tilde{v}_{l,\tau}^{\sigma} q - r_{l, \tau}^{\sigma}}
 \; ,
 \label{eq:Gscalerel2}
 \end{equation} 
and the corresponding single-scale propagator is
\begin{equation}
\dot{ \tilde{G}}_{l, \tau }^{\sigma} ( q , i \epsilon , \alpha ) \approx 
\frac{ \delta ( | q| -1 )}{ 
 i \epsilon - \alpha \tilde{v}_{l, \tau}^{\sigma} q - r_{l, \tau}^{\sigma}}
 \; .
 \label{eq:Gsinglescalerel2}
 \end{equation} 
We recover the vertices of the  momentum
transfer cutoff scheme by taking first the limit $ \tau \rightarrow \infty$.
Of course,
the result of the RG  should be independent of how we reach a 
certain point in two-dimensional cutoff space 
spanned by $\Lambda_l$ and $\Lambda_{\tau}^F$.
Suppose we first fix $\Lambda^F_{\tau} = \Lambda^F_0$
and  perform the reduction of the momentum transfer cutoff 
$\Lambda_0 \rightarrow \Lambda_l$. As a second step we reduce the
fermionic cutoff  
$\Lambda^F \rightarrow 0$. 
On the right-hand side of the flow equation for the
spin-flip susceptibility there are then two additional 
diagrams~\cite{Schuetz05} involving the fermionic single-scale
propagator, which are
shown in Fig.~\ref{fig:flowsuscepfermi}.
For a fixed scale $l$ the corresponding flow equation is
(for simplicity we set $\alpha=1$ and omit the chirality label)
\begin{eqnarray}
\partial_\tau \tilde{\Pi}^{\sigma \bar{\sigma}}_{l,\tau} 
( \bar{Q}) &=& 
- [  \bar{q} \partial_{\bar{q}} + \bar{\epsilon}  \partial_{\bar{\epsilon}} ]
\tilde{\Pi}^{\sigma \bar{\sigma}}_{l,\tau} ( \bar{Q}) 
 +  \dot{{\Pi}}^{\sigma \bar{\sigma}}_{l,\tau} (\bar{Q}) 
  ,
\label{FlowPi}
\end{eqnarray}	
with
\begin{eqnarray}	
	 \dot{\Pi}^{\sigma \bar{\sigma}}_{l,\tau} (\bar{Q}) &=& 
 - {\gamma}_{l,\tau}^2 \int_{Q}  \Bigl[ 
 \dot{\tilde{G}}_{l, \tau}^{\sigma} ({Q}) 
 {\tilde{G}}^{\bar{\sigma}}_{l,\tau} (Q+\bar{Q}+ \sigma \tilde{\Delta}^{\ast}_{\tau}) 
 \nonumber
 \\
 & &  + {\tilde{G}}_{l,\tau}^{\sigma} 
 ({Q}) \dot{\tilde{G}}^{\bar{\sigma}}_{l,\tau}
 (Q+\bar{Q}+ \sigma \tilde{\Delta}_{\tau}^{\ast} ) 
 \Bigr]
 \; ,
\label{DotPi}
\end{eqnarray}
where  $\tilde{\Delta}_{\tau}^{\ast} = \Delta / \Lambda_{\tau}^F$.
The flow equation for the spin-flip vertex is now of the form
\begin{equation}
	\partial_\tau {\gamma}_{l,\tau} = \Theta(  \lambda + l    - \tau )
 C (l,\tau ) 
\; ,
\end{equation}
where  $C ( l , \tau )$ is some function of the flow parameters
and of the running coupling constants, and 
$\lambda = \ln ( \Lambda_0^F / \Lambda_0 )$. 
For simplicity we choose $\Lambda_0 = \Lambda_0^F$, so that $\lambda =0$.
The $\Theta$-function is due to the fact that
the internal loop momenta are restricted by the momentum transfer cutoff
$\Lambda_0$. Obviously, $ \partial_\tau {\gamma}_{l,\tau} =0$ for $\tau >  l$,
so that $\gamma_l \equiv \gamma_{ l , \tau >  l }$ is independent of $\tau$.
Similarly, the
flow equation for $r_{l , \tau }^{\sigma}$ is of the form
\begin{equation}
	\partial_\tau {r}^\sigma_{l,\tau} 
= {r}^{\sigma}_{l, \tau} + \Theta(  l - \tau )   A ( l , \tau )
 \; ,  
\end{equation}
with some other function $A ( l , \tau )$. This
implies $r_{l , \tau }^{\sigma} = e^{ \tau -  l } r^{\sigma}_{ l , \tau = l}$
 for $ \tau >   l$,
so that the flowing Fermi point distance
$\tilde{\Delta}_{ l , \tau } $, defined analogous to 
Eq.~(\ref{eq:tildeDeltaldef}) via
 \begin{equation}
 \tilde{\Delta}_{ l , \tau } =
 \frac{ \Delta }{ \Lambda_\tau^{F} }  
-  ( r^{+}_{ l , \tau   }
 - r^-_{l , \tau } ) 
 \; ,
 \label{eq:Fermidistflow}
 \end{equation}
is for $ \tau >   l$ of the form
 \begin{eqnarray}
 \tilde{\Delta}_{ l , \tau >   l } & = &
 \frac{ \Delta }{ \Lambda_\tau^{F} }  
- e^{( \tau -  l) } ( r^{+}_{ l ,    l  }
 - r^-_{l ,    l  } ) 
 \nonumber
 \\
 & = & 
 e^{\tau -  l } 
 \left[ 
 \frac{ \Delta }{ \Lambda_{  l} }  
-  ( r^{+}_{ l  }
 - r^-_{l   } ) \right]
\nonumber
 \\
 & = & 
 e^{\tau -  l } \tilde{\Delta}_{  l}
\; ,
 \end{eqnarray}
where we have defined $r_l^{\sigma} = r_{ l , \tau =  l}^{\sigma}$ and
$ \tilde{\Delta}_{  l} 
= \Delta / \Lambda_{  l } - ( r_l^{+} - r_l^- )$, see 
Eq.~(\ref{eq:tildeDeltaldef}).
For $ \tau \rightarrow \infty$ 
the solution of Eq.~(\ref{FlowPi}) can therefore
be written as
\begin{eqnarray} 
\tilde{\Pi}^{\sigma \bar{\sigma}}_{l} ( \bar{Q} ) 
&=& \int_0^\infty \!\!\! d\tau \: 
 \dot{\Pi}^{\sigma \bar{\sigma} }_{l, \tau}( \bar{Q} 
e^{\tau-   l } ; \tilde{\Delta}_{l, \tau})
 \nonumber
 \\
&=& \int_0^{  {l}} \!\!\! d\tau \: 
 \dot{\Pi}^{\sigma \bar{\sigma} }_{l, \tau}( \bar{Q} 
e^{\tau-  l } ; \tilde{\Delta}_{l, \tau})
\nonumber
 \\
& + &
  \int_{ {l}}^\infty \!\!\! d\tau \;
 \dot{\Pi}^{\sigma \bar{\sigma} }_{l, \tau}( \bar{Q} 
e^{\tau-  l } ;   \tilde{\Delta}_{  l}  e^{\tau -  l }   )
\label{SolutionPi}
 \; . 
\end{eqnarray}
Using the fact that
in the integral of the last term we may pull out a factor
of $\gamma_l^2$ and
approximating the fermionic Green functions
by Eqs.~(\ref{eq:Gscalerel2}) and
(\ref{eq:Gsinglescalerel2}) with $\tilde{v}^{\sigma}_{ l, \tau}$ and
$r_{l , \tau }$ set equal to zero,
we recover from the last term
 the adiabatic approximation (\ref{eq:Pisfapprox}).
Actually, to calculate the flow of the fermionic self energy, we 
only need the bosonic Green function at $q= \pm 1$. 
For $\tilde{\Delta}_l < 1$ the non-adiabatic 
contribution to the polarization then vanishes,
 leaving us just with the adiabatic part. For larger $\tilde{\Delta}_l$ 
there is a crossover to the more general expression. However,
the leading contribution to the flow of $\tilde{\Delta}_l$ itself stems from the region where the adiabatic approximation is valid.

Finally we note that the adiabatic approximation (\ref{eq:Pisfapprox})
is also consistent with the
flow equation for the spin-flip susceptibility
in the momentum transfer cutoff scheme
shown in Fig.~\ref{fig:flowPi}: taking 
the derivative of the right-hand side 
of Eq.~(\ref{eq:Pisfapprox}) with respect to the flow parameter
and inserting for the derivative of the fermionic self-energy
its flow equation (\ref{eq:RGselfvertex}), we find
that the right-hand side can be written in terms of
a symmetrized closed fermion loop with four bosonic legs and
renormalized propagators. This corresponds to
the adiabatic approximation
for the vertex $\Gamma^{(0,4)}_{\Lambda}$
which according to Fig.~\ref{fig:flowPi} drives
the flow  of  the spin-flip susceptibility.

%\vspace{-4mm}

\end{document}